\documentclass[prd,showpacs,amsmath,amssymb,nofootinbib,superscriptaddress]{revtex4}
\usepackage{array,mathtools,amssymb,booktabs}
\usepackage[english]{layout}
\usepackage[english]{babel}
\usepackage{bm}
\usepackage{graphicx}

\newcommand{\Od}{{\cal O}}
\newcommand{\IZ}{{\Bbb Z}}
\newcommand{\IR}{{\Bbb R}}
\newcommand{\im}{\mbox{Im}\,}
\newcommand{\re}{\mbox{Re}\,}
\newcommand{\tintq}{T\sum_n \int\frac{d^3 \vec{q}}{(2\pi)^3}}
\newcommand{\modp}{\vert \vec{p} \vert}
\newcommand{\modq}{\vert \vec{q} \vert}
\newcommand{\pif}{\bm{\pi}}

\begin{document}

\title{Large-$N$ pion scattering at finite temperature: the $f_0(500)$ and chiral restoration}
\author{Santiago Cort\'es}
\email{js.cortes125@uniandes.edu.co}
\affiliation{Departamento de F\'{\i}sica, Univ. de Los  Andes, Bogot\'a, Colombia.}
\author{A.~G\'omez Nicola}
\email{gomez@ucm.es}
\affiliation{Departamento de F\'{\i}sica
Te\'orica II. Univ. Complutense. 28040 Madrid. Spain.}
\author{John Morales}
\email{jmoralesa@unal.edu.co}
\affiliation{Departamento de F\'{\i}sica, Univ. Nacional de Colombia, Bogot\'a, Colombia.}
\date{\today}

\begin{abstract}
We consider the $O(N+1)/O(N)$ Non-Linear Sigma Model for large $N$ as an effective theory for low-energy QCD at finite temperature $T$, in the chiral limit. At $T=0$ this formulation provides a good description of scattering data in the scalar channel and generates dynamically the $f_0(500)$ pole, the pole position lying within experimental determinations.  Previous $T=0$ results with this model are updated using newer analysis of pion scattering data. We calculate the pion scattering amplitude at finite $T$  and show that it satisfies exactly thermal unitarity, which had been assumed but not formally proven in previous works. We discuss the main differences with the $T=0$ result and we show that one can define a proper renormalization scheme with $T=0$ counterterms such that the renormalized amplitude can be chosen to depend only on a few parameters. Next, we analyze the behaviour of the $f_0(500)$ pole at finite $T$, which is consistent with chiral symmetry restoration when the scalar susceptibility is saturated by the $f_0(500)$ state, in a second-order transition scenario  and in accordance with lattice and theoretical analysis. 

\vspace*{0.5cm}
\end{abstract}

\maketitle

\section{Introduction}

The study of hadronic properties at finite temperature $T$ is one of the theoretical ingredients needed to understand the behaviour of  matter created in Relativistic Heavy Ion Collision experiments, such as those  in RHIC and LHC (ALICE). In particular, the QCD transition involving chiral symmetry restoration and deconfinement plays a crucial role, as it is clear from the many recent advances of lattice groups in the study of the phase diagram and other thermodynamical properties \cite{Aoki:2009sc,Borsanyi:2010bp,Bazavov:2011nk,Buchoff:2013nra,Bhattacharya:2014ara}.

For vanishing baryon chemical potential, the QCD transition is a crossover for  2+1 flavours with physical quark masses, the transition temperature being about  $T_c\sim$ 150 -160 MeV. In the chiral limit (vanishing light quark masses for fixed strange mass) it is believed to become  a second-order phase transition belonging to the universality class of the $O(4)$ model \cite{Pisarski:1983ms,Berges:2000ew}. Lattice simulations also support this fact.   Actually, in \cite{Bazavov:2011nk,Ejiri:2009ac} it is shown that the lattice results are compatible with a $O(N)$-like  restoration pattern in the chiral limit and for physical masses, by studying the scaling of different thermodynamical observables near $T_c$. The expected reduction in the transition temperature from the physical mass case to the chiral limit one based on those analysis is about 15-20$\%$, although subject to many lattice uncertainties \cite{Bazavov:2011nk} .

From the theoretical side, it  is important to provide solid analysis of this chiral restoration pattern based on effective theories, given the limitations of perturbative QCD at those temperature scales. A simple model realization  was historically the Linear Sigma Model (LSM) based on $O(4)\rightarrow O(3)$ spontaneous symmetry breaking  \cite{Pisarski:1983ms,Hatsuda:1985eb, Bochkarev:1995gi}, where the $\sigma$-component of the $O(4)$ field acquires a thermal vacuum expectation value and mass, both of them vanish at the transition in the chiral limit, and  $\pi-\sigma$ mesons degenerate as chiral partners.  However, such a simple description is nowadays in conflict with observations: on the one hand, the $\sigma/f_0(500)$  broad resonance produced in pion-pion scattering and listed in the PDG  \cite{Agashe:2014kda} is not compatible with a particle-like state (see \cite{Pelaez:2015qba} for a recent review).  On the other hand, to reproduce consistently pion data,  the LSM requires working in a strong coupling regime, invalidating the perturbative description. Nevertheless, it is clear that the $\sigma/f_{0}(500)$ state must play an important role in chiral restoration, since it shares the quantum numbers of the QCD vacuum. Chiral symmetry restoration has also been studied within QCD inspired models like the Nambu-Jona-Lasinio or Gross-Neveu ones \cite{Bernard:1987ir,Blaizot:2002nh,Buballa:2003qv,master}.

A systematic and model-independent framework that takes into account the relevant light meson degrees of freedom and their interactions is Chiral Perturbation Theory (ChPT) \cite{Weinberg:1978kz,Gasser:1983yg}. The effective ChPT Lagrangian is constructed as a derivative and mass expansion  ${\mathcal L}={\mathcal L}_{p^2}+{\mathcal L}_{p^4}+\dots$, where $p$ denotes generically a meson energy scale compared to the chiral scale $\Lambda_{\chi}\sim$ 1 GeV. The lowest order Lagrangian ${\mathcal L}_{p^2}$ is the Non-linear Sigma Model (NLSM). The use of energy expansions in chiral effective theories is also justified at finite temperature to describe Heavy Ion Physics. Pions are actually the most copiously produced particles after a Heavy Ion Collision and most of their properties from hadronization to thermal freeze-out can be reasonably described within the temperature range where these theories are applicable. Thus, the chiral restoring behaviour in terms of the quark condensate is qualitatively obtained within ChPT \cite{Gasser:1986vb,Gerber:1988tt}.  Moreover,  the introduction of realistic pion interactions by demanding unitarity through the Inverse Amplitude Method (IAM) \cite{Dobado:1996ps} extended at finite $T$ \cite{GomezNicola:2002tn,Dobado:2002xf}  improves ChPT, providing a more accurate description of several effects of interest in a Heavy-Ion environment, such as thermal resonances, transport coefficients and electromagnetic corrections \cite{FernandezFraile:2009mi,Nicola:2014eda}. This approach  also provides a novel understanding of the role of the $\sigma/f_0(500)$ broad resonant state in chiral symmetry restoration, without having to deal with the typical LSM drawbacks. Thus, the unitarized $\pi\pi$ scattering amplitude within ChPT at finite temperature develops a $I=J=0$ thermal pole at $s_p= \left[M_p(T)-i \Gamma_p(T)/2\right]^2$,  which for $T=0$ corresponds to the PDG state, and  whose trajectory in the complex plane as $T$ varies shows some interesting features: the sudden drop of $M_p (T)$ towards the two-pion threshold can be interpreted in terms of chiral symmetry restoration, as opposed for instance to the  $I=J=1$ $\rho$-channel where the mass drop is much softer. In addition, it has been recently shown \cite{Nicola:2013vma} that the scalar susceptibility  saturated with this $\sigma$-like state, with squared mass $M_S^2=M_p^2-\Gamma_p^2/4$, develops a maximum near $T_{c}$ compatible with lattice data, unlike the pure ChPT prediction which is monotonically increasing. Moreover, chiral partners in the scalar-pseudoscalar sector are understood through degeneration of correlators and susceptibilities \cite{Nicola:2013vma}, something which is also directly seen in lattice data \cite{Buchoff:2013nra,Bhattacharya:2014ara}. The role of the $f_0(500)$ state for chiral restoration could become more complicated if its possible tetraquark component is also considered at finite temperature  \cite{Heinz:2008cv}.

A crucial step in the unitarized approach is the condition of exact thermal unitarity for the partial waves, with a thermal space factor modified by the Bose-Einstein distribution function. This condition holds perturbatively in ChPT \cite{GomezNicola:2002tn} and the unitarized amplitude is constructed by requiring thermal unitarity to all orders, based on the physical collision processes occurring in the thermal bath  \cite{Dobado:2002xf,GomezNicola:2002an}. However, it is important to emphasize that thermal unitarity for the full amplitude was not formally proven in those works; in fact, that will be one of the relevant issues discussed in the present work.

Although the approaches based on effective theories in terms of the lightest mesons provide a good description of the physics involved, especially in what concerns the effect of the lightest resonances (as discussed above), a more accurate treatment near $T_{c}$ would require including heavier degrees of freedom. That is for instance the framework of the Hadron Resonance Gas, which describes the system through the statistical ensemble of all free states thermally available, and where corrections due to interactions and lattice masses can be also accounted for \cite{Andronic:2008gu,Huovinen:2009yb}.  Effective chiral models including explicitly  vector and axial-vector resonances have also been successfully used to depict several hadron thermal properties relevant for observables such as the dilepton and photon spectra and $\rho-a_1$ mixing/degeneration at the chiral transition \cite{Rapp:1999ej,Turbide:2003si}.

In this work, we will consider an alternative approach to the thermal pion scattering amplitude, namely the limit of large number of Nambu-Goldstone bosons $N$, or in other words, large number of light flavours with no strangeness, as treated before at $T=0$ in \cite{Dobado:1992jg,Dobado:1994fd}.  Previous large-$N$ analysis at $T\neq 0$ can be found in \cite{Bochkarev:1995gi,Jeon:1996gn,Andersen:2004ae}.  Within this framework, the lowest order chiral effective Lagrangian for low-energy QCD will be the $O(N+1)/O(N)$ NLSM, whose corresponding symmetry breaking pattern is $O(N+1)\rightarrow O(N)$. As we have just commented, the latter is believed to take place in chiral symmetry restoration for $N=3$, since $O(4)$ and $O(3)$ are respectively isomorphic to the isospin groups $SU_L(2)\otimes SU_R(2)$ and $SU_V(2)$. This technique has the advantage of allowing for a partial resummation of the scattering amplitude preserving many physical properties such as unitarity and the dynamical generation of the $f_0(500)$ pole, which will help us to shed more light on the chiral restoring issues discussed before. We will work in the chiral limit, since it simplifies considerably the analysis, besides enhancing chiral-restoring effects, as explained above. At this point, it is important to remark that massless pions remain massless at finite temperature \cite{Gasser:1986vb, Gerber:1988tt}, unlike many other instances in thermal field theory where elementary massless excitations acquire mass in the thermal bath, like  high-$T$ fermions \cite{Weldon:1982bn}, gauge fields \cite{lebellac,Kapusta:2006pm} including large-$N_f$ analysis \cite{Blaizot:2005fd} or when electromagnetic corrections are switched on \cite{Nicola:2014eda}. 

The study of the large-$N$ approach  in low-energy QCD implies a simplification of the pion dynamics \cite{Dobado:1992jg,Dobado:1994fd} without changing essential features such as analyticity, unitarity and the low-energy behaviour for pion scattering. This is fully accomplished when working in the functional  formalism of the theory \cite{Gerstein:1971fm,Appelquist:1980ae,Dobado:1992jg,Dobado:1994fd,dobadobook} so that the Lagrangian is built as $O(N+1)/O(N)=S^N$ covariant and $O(N+1)$ invariant (in the chiral limit).  Furthermore, as we will see in detail here, this approach will allow to describe consistently the $f_0(500)$ state through its pole in the second Riemann sheet, where the parameters of the model are fitted to pion-pion scattering phase shift data. Thanks to the fact that the model is exactly unitary, we will be  granted to go beyond the standard perturbative ChPT description for the scattering \cite{Gasser:1983yg,bijnens}, as a complementary description of  unitarization methods such as the IAM. 

An additional observation that makes this approach suitable for studying chiral restoring effects is that in order to reproduce correctly the $f_{0}(500)$ pole (linked to chiral restoration as mentioned before), the dominant contributions to $\pi\pi$ scattering are the loop diagrams  from the leading order chiral Lagrangian, rather than the particular form  of higher order terms needed to renormalize the amplitude \cite{Dobado:1996ps,Oller:1997ng,Pelaez:2006nj}. Thus, the large-$N$ limit framework provides a resummation of  the dominant loop contributions needed  to maintain exact unitarity, so that the scalar pole can be  correctly described.

With the above motivations kept in mind,  we will analyze in this work elastic pion-pion scattering  at finite temperature within the large-$N$ $O(N+1)/O(N)$ model in the chiral limit. We will show that at $T=0$ one gets reasonable values for the $f_{0}(500)$ pole from a fit to experimental data of a two-parameter partial wave. The extension to $T\neq 0$ includes a formal discussion of the renormalizability of the model, which as expected can be carried out in terms of $T=0$ counterterms, although with important subtleties to be taken into account. The important feature of exact unitarity is demonstrated, including thermal corrections, something that allows us to define the second-sheet pole. Having fixed the $T=0$ pole position, its $T$ dependence is obtained and it is shown that the results are compatible with a second-order chiral restoring phase transition, consistently with previous determinations and  lattice data.

The paper is organized as follows: In section \ref{sec:scattering} we introduce our large $N$ approach within the framework of the massless NLSM and work out the diagrammatic expansion for pion scattering, both at zero and at finite temperature; section \ref{sec:ren} is devoted to explain the renormalization procedure, for which  technical details are relegated to Appendix \ref{app:ren}. In section \ref{sec:pw} we perform the analysis of the $I=J=0$ partial wave, providing a fit to $T=0$ data and showing that the large-$N$ amplitude satisfies exactly unitarity at zero and finite temperature. The latter grants us to define the Riemann second-sheet pole corresponding to the $f_0(500)$ state, which we study in detail in section \ref{sec:pole}, paying special attention to its thermal evolution and the connection with chiral symmetry restoration. Our conclusions are presented in section \ref{sec:conc}.

\section{Pion scattering amplitude in the $O(N+1)/O(N)$ NLSM}
\label{sec:scattering}

\subsection{Diagrammatics at Zero Temperature} 

In a theory with spontaneous symmetry breaking  $O(N+1)\rightarrow O(N)$, the coset space where the Nambu-Goldstone Bosons (NGB) are defined is the $N$-dimensional sphere $S^{N}=O(N+1)/O(N)$. In such a theory, the most general $O(N+1)$-invariant and $S^{N}$ covariant Lagrangian in the chiral limit can be obtained as a derivative expansion of the NGB modes, whose lowest order expression is given by the NLSM \cite{Dobado:1992jg,Dobado:1994fd,dobadobook}:

\begin{equation}
\mathcal{L}_{NLSM}=\frac{1}{2}g_{ab}(\pif)\partial_{\mu}\pif^{a}\partial^{\mu}\pif^{b},
\label{NLSM}
\end{equation}
where $g_{ab}(\pif)$ is a metric in the $S^{N}$ manifold which is parametrized in the NGB $\pif_a$ coordinates as

\begin{equation}
g_{ab}(\pif)=\delta_{ab}+\frac{1}{NF^{2}}\frac{\pif_{a}\pif_{b}}{1-\pif^2/NF^{2}},
\label{metric}
\end{equation}
with $\pif^2=\displaystyle\sum_{a=1}^{N}\pif_{a}\pif^{a}$. 

As explained in \cite{dobadobook}, the advantage of choosing a $S^{N}$ covariant formalism is that we can easily construct $O(N+1)$ invariant Lagrangians of higher order by properly contracting indices with the $g_{ab}$ metric. In addition, this formalism ensures the independence of the Green functions on NGB field reparametrizations, i.e., when changing coordinates in $S^N$. The covariance of the quantum model is  guaranteed as long as we work in the Dimensional Regularization  (DR) scheme with $D=4-\epsilon$, since the metric factor appearing in the NGB quantum measure $(\mathcal{D}\pif\sqrt{g})$, with $g$ the metric determinant, amounts to add to the Lagrangian a term proportional to $\delta^{D}(0)$ \cite{Gerstein:1971fm,Appelquist:1980ae, zjbook} which vanishes in DR \cite{Leibbrandt:1975dj}.

Although there is no need to invoke the LSM in the above construction, one can understand the NLSM  as the kinetic part of the LSM when the vacuum constraint $\pif^2+\sigma^{2}=NF^{2}$ is imposed, where $v=\sqrt{N}F$ is the vacuum expectation value acquired by the $\sigma$ field at tree level. In that way, it is easy to understand the $N$-scaling of the $NF^2$ constant, where $F_{\pi}^{\,2}=NF^{2}$ is the pion decay constant at this order for the usual  $N=3$ case \cite{Gasser:1983yg}.

The Lagrangian \eqref{NLSM} provides the standard kinetic term for the NGB fields and along with it, an infinite set of self-interaction terms with an arbitrary even number of NGB. These interactions are obtained when the metric (\ref{metric}) is expanded and written as a function of two field derivatives and powers of the pion field. Hence, to obtain  the relevant Feynman rules for pion scattering, we  can write the Lagrangian (\ref{NLSM}) as follows:

\begin{equation}
\mathcal{L}_{NLSM}=\frac{1}{2}\partial_{\mu}\pif_{a}\partial^{\mu}\pif^{a}+\frac{1}{8NF^{2}}(\partial_{\mu}\pif^2)^{2}\left[1+\frac{\pif^2}{NF^{2}}+\left(\frac{\pif^2}{NF^{2}}\right)^{2}+\cdots\right].
\label{lagrangian1}
\end{equation}

\begin{figure}
\centering
\includegraphics[scale=1]{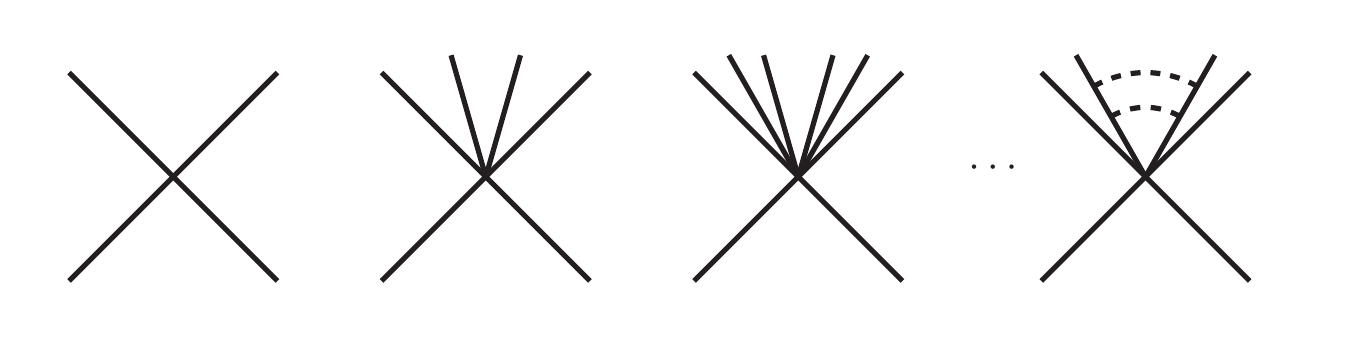}
\caption{Feynman rules and diagrams at tree level, where dashed lines correspond to multiple pairs of pion lines.}
\label{treel}
\end{figure}

The first term in the latter expansion gives the standard kinetic Lagrangian and the rest are the $2n$-vertices with $n\geq 2$, represented in Fig. \ref{treel}. The Feynman rule in momentum space for each vertex is $\frac{(p_A+p_B)\cdot(p_C+p_D)}{(NF^2)^{n-1}}\delta_{AB}\delta_{CD}\dots$ with $A,B$ and $C,D$ the isospin indices of all possible choices of four different lines in the diagram and where the dots indicate the rest of pair contractions, i.e., all products $\delta_{ij}$ with $ij$ the isospin indices of pairs different from $AB$ and $CD$. With these rules, we proceed to calculate the $\pi^{a}\pi^{b}\rightarrow \pi^{c}\pi^{d}$ scattering that, as customary, is parametrized as:

\begin{equation}
T_{abcd}(s,t,u)=\delta_{ab}\delta_{cd}A(s,t,u)+\delta_{ac}\delta_{bd}A(t,s,u)+\delta_{ad}\delta_{bc}A(u,t,s).
\label{generalT}
\end{equation}
Here, $s,t,u$ are the  Mandelstam variables $s=(p_a+p_b)^2=(p_c+p_d)^2$, $ t=(p_a-p_c)^2$, $u=(p_a-p_d)^2$ and use of isospin and crossing symmetry has been made to parametrize the amplitude. We will denote $p=p_a+p_b$.

\begin{figure}
\centering
\includegraphics[scale=1]{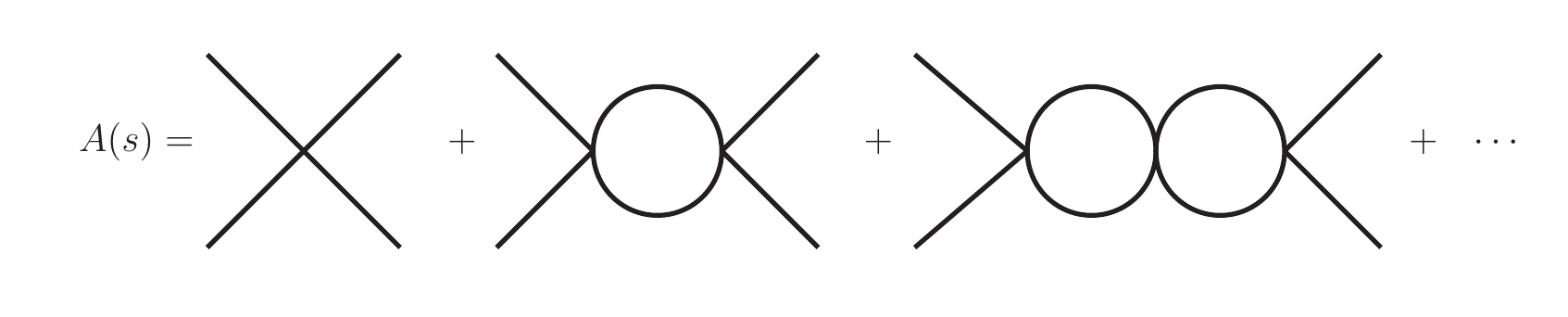}
\caption{Zero-Temperature scattering amplitude.}
\label{4pioneff}
\end{figure}

The dominant contribution  to $A(s,t,u)$ in the large-$N$ limit comes from the diagrams showed in Fig. \ref{4pioneff}. Thus, when considering  diagrams of arbitrary loop order with just the four pion vertex in Fig.\ref{treel}, the dominant contribution of isospin flux is that where the 
pairs of lines are chosen as $(p_A+p_B)\cdot(p_C+p_D)=(q_1+p-q_1)\cdot(q_2+p-q_2)=s$ for an internal vertex, i.e., with no attachments to any external line ($q_1$ and $q_2$ are the four-momenta of the loops attached to that vertex) and 
$(p_A+p_B)\cdot(p_C+p_D)=p\cdot(q_1+p-q_1)=s$ for the external ones. In that way, an additional factor of $N$ is generated for every pair of vertices between a given loop, coming from a contraction  $\delta_{ef}\delta^{ef}=N$, where $e$ ($f$) is one of the two free indices in the first (second) vertex. The result is a net factor $s/(NF^{2})$ for every vertex and an additional $N$ for every loop, so that the resulting amplitude is $\Od(1/N)$. Other loop contributions are subdominant according to this counting.

For those vertices in Fig.\ref{treel} with more than four pions, the only way to compensate the additional $(1/N)^{n-1}$ factors of the $2n$-pion vertex is to close $2n-4$ of them in tadpole-like contributions, giving rise to a $N^{n-2}$ factor, so that this contribution would count the same $1/N$ as the four-pion vertex. Those tadpole insertions correspond to fields sharing the same isospin index with no derivatives, since $\partial_\mu G(x)\vert_{x=0}=0$, with $G(x)$ the free pion propagator. At $T=0$, they actually vanish, since the tadpole contribution $G_{T=0}(x=0)=0$ in the chiral limit.  That will be not the case at $T\neq0$, as we discuss in section \ref{sec:diagft} and will become one of the main novelties of the present calculation. Finally, note also that in the chiral limit, the pion propagator is not corrected by loop effects and hence needs no renormalization \cite{Appelquist:1980ae}. For instance,   a tadpole correction to the self-energy would require contracting two pion lines with the same isospin index to produce a $N$ factor, but that gives $\partial_\mu G(x)\vert_{x=0}=0$, and other contributions are non-dominant with respect  to the tree level propagator, which is $\Od(1)$. 

From the latter considerations, and after including the proper combinatoric factors, the $A(s,t,u)$ function in  (\ref{generalT}) depends only on $s$ to leading order in $1/N$,  which is actually one of the main simplifications of this approach, and is given by

\begin{equation}
A(s)=\frac{s}{NF^{2}}\left[1+\frac{1}{2}\frac{1}{F^2}s\,J(s)+\frac{1}{4}\frac{1}{F^4}s^2\,J^2(s)+\cdots\right]=
\frac{s}{NF^{2}}\sum_{k=0}^\infty \left[\frac{s J(s)}{2F^2}\right]^k=
\frac{s}{NF^{2}[1-s\,J(s)/2F^{2}]},
\label{zerotamp}
\end{equation}
where $J(s)$ is the usual logarithmically-divergent loop integral that in the DR scheme reads \cite{Gasser:1983yg}

\begin{equation}
J(s)=-i\int \frac{d^D q}{(2\pi)^D}\frac{1}{q^2}\frac{1}{(p-q)^2}=J_\epsilon(\mu)+\frac{1}{16\pi^2}\ln \left(\frac{\mu^{2}}{-s}\right)
\label{Js}
\end{equation}
and $J_\epsilon (\mu)$ contains the divergent part:

\begin{equation}
J_\epsilon(\mu)=-2\lambda(\mu)+\frac{1}{16\pi^2}=\frac{1}{16\pi^2}\left[\frac{2}{\epsilon}+\ln (4\pi)-\gamma+2-\ln\mu^2\right]+\Od(\epsilon).
\label{Jepsilon}
\end{equation}

Here $\lambda(\mu)=\frac{\Gamma(1-D/2)}{2(4\pi)^{D/2}}\mu^{D-4}$, $\gamma$ is Euler's constant and $\mu$ is the renormalization scale.  Note that we follow the convention in \cite{Gasser:1983yg} to define the pole contribution $\lambda(\mu)$ but we include the $1/(16\pi^2)$ contribution in the divergent part, unlike in \cite{Gasser:1983yg}, in order to compare easily with previous large-$N$ works \cite{Dobado:1992jg,Dobado:1994fd}. 

We recall that for $s\in\IR$ and $s\geq0$ (i.e., above the two-pion threshold which in the chiral limit is at $s=0$) we can easily obtain the imaginary parts of the loop integral as the usual unitarity cut contribution (see section \ref{sec:unit}):

\begin{equation}
\im J(s+i0^+)=  \frac{1}{16\pi}                 \qquad (s>0)
\label{imJ0}
\end{equation}
while $\im J(s)=0$ for $s<0$.

In section \ref{sec:ren} we will discuss the renormalization procedure implemented to absorb the divergent part (\ref{Jepsilon}), but before that, let us explain the main distinctive features of the $T\neq 0$ calculation.

\subsection{Diagrammatics at Finite Temperature} 
\label{sec:diagft}

We will work within the imaginary-time formalism of Thermal Field Theory \cite{lebellac,Kapusta:2006pm}  so that the thermal scattering amplitude is understood as the analytic continuation to continuous energies of the corresponding four-point Green function  after performing the loop Matsubara sums and applying the LSZ standard reduction formula for $T=0$ asymptotic states  \cite{GomezNicola:2002tn}.

Comparing with the analysis performed in the previous section, the first observation is that the absence of renormalization for the pion propagator remains the same at $T\neq 0$, so pions do not acquire effective thermal masses, following the same diagrammatic argument as before. However, there is an important difference with the $T=0$ case and is that the tadpole contribution is now different from zero in the chiral limit, namely \cite{lebellac}: 

\begin{equation}
G_{T}(x=0)\equiv I_\beta=\tintq \frac{1}{\omega_n^2+\modq^2}=\frac{T^2}{12},
\label{tadpole}
\end{equation} 
with the Matsubara frequencies $\omega_k=2\pi kT$, $k\in\IZ$.

\begin{figure}
\centering
\includegraphics[scale=1]{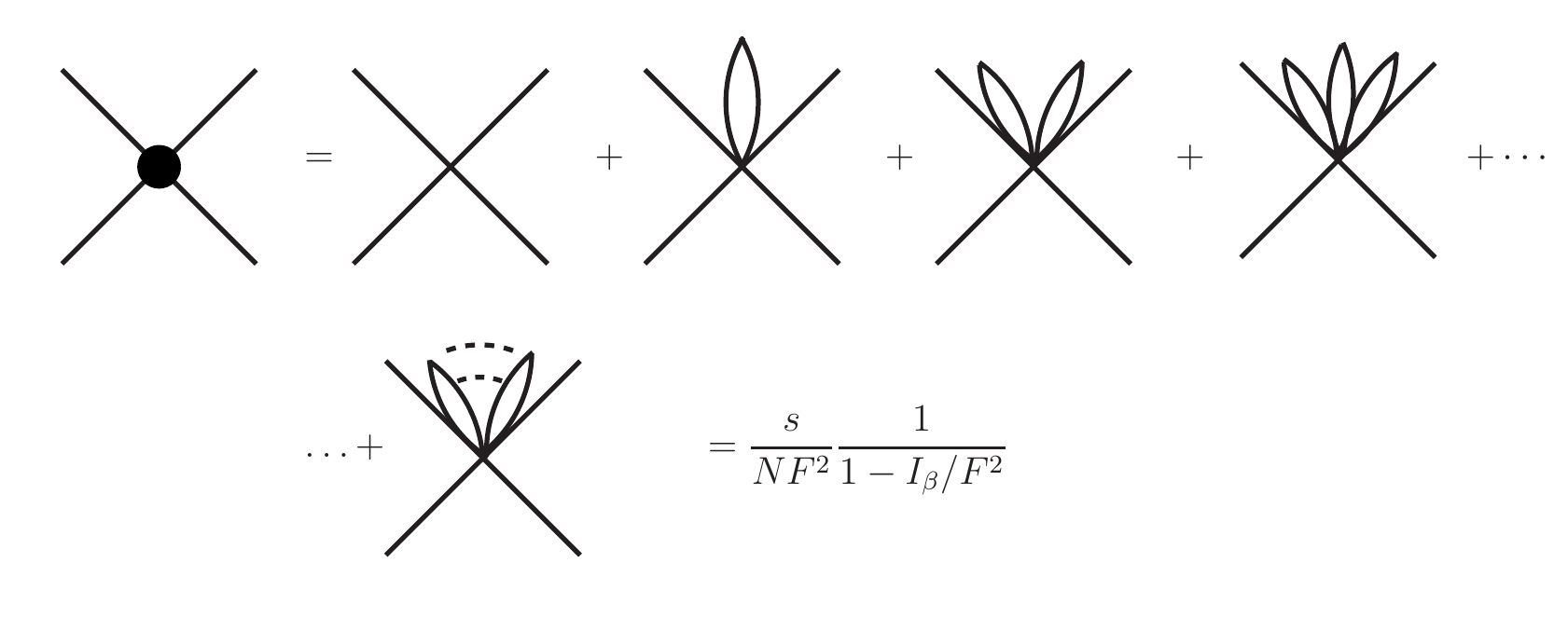}
\caption{Construction of the effective thermal tadpole vertex, where dashed lines correspond to multiple tadpole insertions.}
\label{EffVertx}
\end{figure}

\begin{figure}
\centering
\includegraphics[scale=1]{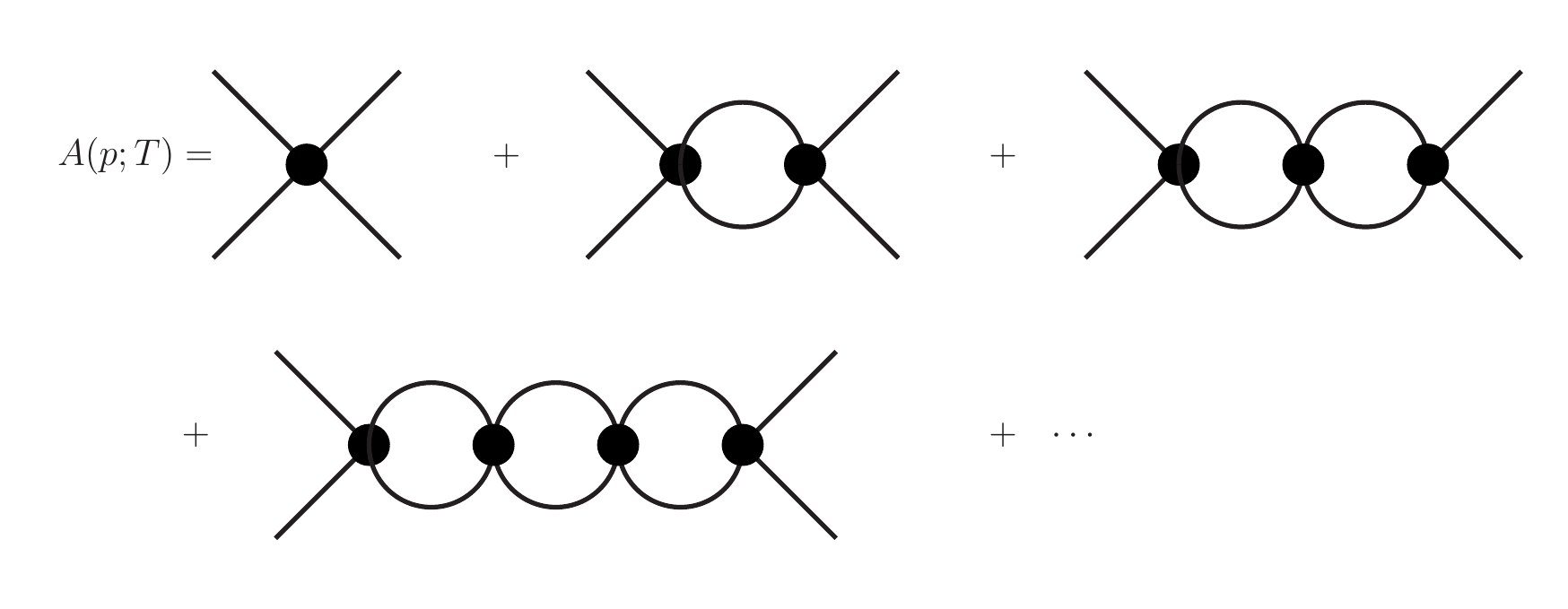}
\caption{Finite-Temperature scattering amplitude.}
\label{Finitetamp}
\end{figure}

This means that from now on, the diagrams coming from closing pairs of extra pion lines in the vertices with 6 or more legs in Fig.\ref{treel} have to be considered. To accomplish this in an efficient way, we construct the effective thermal tadpole vertex given in Fig.\ref{EffVertx}, and we rebuild the scattering amplitude to all perturbative orders with the associated Feynman rule of the thermal vertex, something that we show schematically in Fig.\ref{Finitetamp}.

In addition, we have to take into account that the loop integral is also $T$-dependent, so that  the thermal amplitude to leading order in $1/N$ is given by 

\begin{equation}
A(p_{0},\modp;T)\equiv A(p;T)=\frac{s}{NF^{2}}\frac{f(I_{\beta})}{1-\frac{s}{2F^{2}}f(I_{\beta})J(p_0,\modp;T)},
\label{thermalampbare}
\end{equation}
which depends now separately on the space and time components of $p$ due to the loss of Lorentz covariance in the thermal bath. The vertex function reads

\begin{equation}
f(I_{\beta})=\frac{1}{1-I_\beta/F^2},
\end{equation}
and the finite-$T$ loop integral $J(p_0,\modp;T)$ is the analytic continuation of the external Matsubara frequency $i\omega_m\rightarrow p_0+i0^+$ of

\begin{equation}
J(i\omega_m,\modp;T)=\tintq \frac{1}{\omega_n^2+\vert\vec{q}\vert^2}\frac{1}{(\omega_n-\omega_m)^2+\vert \vec{p}-\vec{q}\vert^2},
\label{JT}
\end{equation}
 where $J(p_0,\modp;T=0)=J(s)$ in  (\ref{Js}).

Explicit expressions for the above $J$ integral for arbitrary three-momentum $\vec{p}$ can be found for instance in \cite{Nicola:2014eda}. Its UV divergent part is the same as for $T=0$, since Bose-Einstein factors regulate exponentially the UV behaviour, so that we will write in general

\begin{equation}
J(p;T)=J_\epsilon (\mu)+ J_{fin}(p;T;\mu),
\label{JTsepar}
\end{equation}
with $J_\epsilon (\mu)$ in \eqref{Jepsilon} and $J_{fin}(p;T;\mu)$ finite and whose scale dependence is contained only in the  $T=0$ part, namely $J_{fin}(p;T=0;\mu)=\frac{1}{16\pi^2}\ln(-\mu^{2}/s)$. 

In this work we will be interested only in calculations in the center of mass frame (corresponding to $\vec{p}=\vec{0}$), where partial waves are defined (see section \ref{sec:pw}) and moreover, we have \cite{GomezNicola:2002tn}

\begin{align}
J_{fin}(p_0,\vec{0};T;\mu)=\frac{1}{16\pi^2}\ln \left(\frac{\mu^{2}}{-s}\right)
+\delta J(s;T); \label{Jthermal1}\\
\delta J(s;T)=\frac{1}{\pi^2}\int_0^\infty dy \frac{y \ n_B(y)}{4y^2-s},
\label{Jthermal2}
\end{align}
where $s=p_{0}^{\,2}$ and $n_B(x)$ is the usual Bose-Einstein distribution function \[n_B(x)=\frac{1}{\exp(x/T)-1}.\]

Note that $\delta J(s;T)$ is UV finite ($y\rightarrow\infty$) thanks to the $n_B(y)$ term; besides, we can easily separate the real and imaginary parts of $\delta J(s;T)$ for $s\in\IR$ and $s>0$ by isolating the pole contribution at $y=\sqrt{s}/2=\vert p_0\vert/2$ in the integrand in (\ref{Jthermal2}) as

\begin{align}
\re\delta J(s;T)= {\cal P} \frac{1}{\pi^2}\int_0^\infty dy \frac{y \ n_B(y)}{4y^2-s},\\
\im\delta J(s+i0^+;T)=\frac{1}{8\pi}n_B(\sqrt{s}/2) \qquad (s>0),
\label{imJther}
\end{align}
while for $s<0$ there is no pole in the integrand in (\ref{Jthermal2}).

Finally, we recall that for $s\neq 0$, $\delta J(s;T)$ is  IR finite ($y\rightarrow 0^+$) while for $s\rightarrow 0^+$, it diverges as $\delta J(s;T)\sim s^{-1/2}$ so that $s\delta J(s;T)$ (as it appears in the thermal  amplitude (\ref{thermalampbare})) remains finite (and vanishing) in that limit. 

\section{Renormalization of the scattering amplitude}
\label{sec:ren}

We will first review  the main points regarding the renormalization of the model in the $T=0$ case, as discussed in \cite{Dobado:1992jg,Dobado:1994fd}. The scattering amplitude can be renormalized by choosing an appropriate (infinite) set of counterterm Lagrangians of higher orders in derivatives and summing their contribution to the amplitude to all orders. The philosophy behind this approach is to include only those Lagrangians needed to obtain a renormalized amplitude to leading order in $1/N$, although from the symmetry arguments explained above, many other operator structures are possible. Consistently, we can consider formally the couplings, or low-energy constants (LEC) of those additional Lagrangians to be suppressed in the $1/N$ counting. In the conventional ChPT approach \cite{Gasser:1983yg,bijnens} all possible terms are included to a given order in the derivative/momentum expansion and then the LEC are fixed with experimental data, although the predictions are limited to low energies. The energy applicability range can be enlarged when  additional conditions such as unitarization are implemented, and then the LEC can take in general different values from the perturbative ones. Here we are considering a partial resummation of the series for the amplitude, namely the leading $1/N$ contribution, which in the end can be given in a finite form that depends only on a few parameters, to be fixed to experimental data.  Nevertheless,  there will be some important subtleties to be taken into account in the $T\neq 0$ case, as we will explain below, in order to ensure a renormalized amplitude  with a $T=0$  renormalization scheme. We will discuss the main results in this section, while additional  details are given in Appendix \ref{app:ren}. 
  
Let us consider, for instance, the possible counterterms Lagrangians to fourth-order in derivatives. It is clear that one of them satisfying the symmetry constraints would be just proportional to ${\mathcal L}_{NLSM}^2$:

\begin{equation}
{\cal L}_1=\frac{g_1}{2NF^4}\left[g_{ab}(\pif)\partial_{\mu}\pif^{a}\partial^{\mu}\pif^{b}\right]^2=\frac{g_1}{2NF^4}\left[\left(\partial_\mu\pif^a\partial^\mu\pif_a\right)^2+\Od\left(\frac{\pif^6}{N}\right)\right],
\label{L4}
\end{equation}
where the normalization proportional to the bare coupling $g_1$ has been  conveniently chosen. At this order there is another term allowed, i.e. ${\cal L}'_1\sim\left[g_{ab}(\pif)\partial_{\mu}\pif^{a}\partial_{\nu}\pif^{b}\right]\left[g_{cd}(\pif)\partial^{\mu}\pif^{c}\partial^{\nu}\pif^{d}\right]$. The LEC multiplying the two allowed terms are the counterpart of the $l_1,l_2$ constants of ChPT \cite{Gasser:1983yg}. 

The main result at $T=0$  is that the scattering amplitude can be rendered finite by a set of infinite counterterm Lagrangians which to $\Od(\pif^4)$ have the form \cite{Dobado:1992jg}

\begin{equation}
{\cal L}_k=(-1)^{k+1}2^{k-2}\frac{g_k}{NF^{2(k+1)}}\left[\partial_{\mu_1}\partial_{\mu_2}\cdots \partial_{\mu_k}\pif^a\partial^{\mu_1}\partial^{\mu_2}\cdots \partial^{\mu_k}\pif_a\partial^\nu \pif^b\partial_\nu\pif_b+\Od\left(\frac{\pif^6}{N}\right)\right].
\label{Lk}
\end{equation}
This reduces to \eqref{L4} for $k=1$. It is indeed always possible to find at each order an adequate contraction with the $g_{ab}$ metric that gives rise to the terms (\ref{Lk}), for instance $g_{ab}(\pif)g_{cd}(\pif)\partial_{\mu_1}\partial_{\mu_2}\cdots \partial_{\mu_k}\pif^a\partial^{\mu_1}\partial^{\mu_2}\cdots \partial^{\mu_k}\pif^b\partial^\nu \pif^c\partial_\nu\pif^d$.  Consistently, the LEC multiplying other possible terms such as ${\cal L}'_{1}$ can be considered formally suppressed in the $1/N$ counting. 

\begin{figure}
\centering
\includegraphics[scale=1]{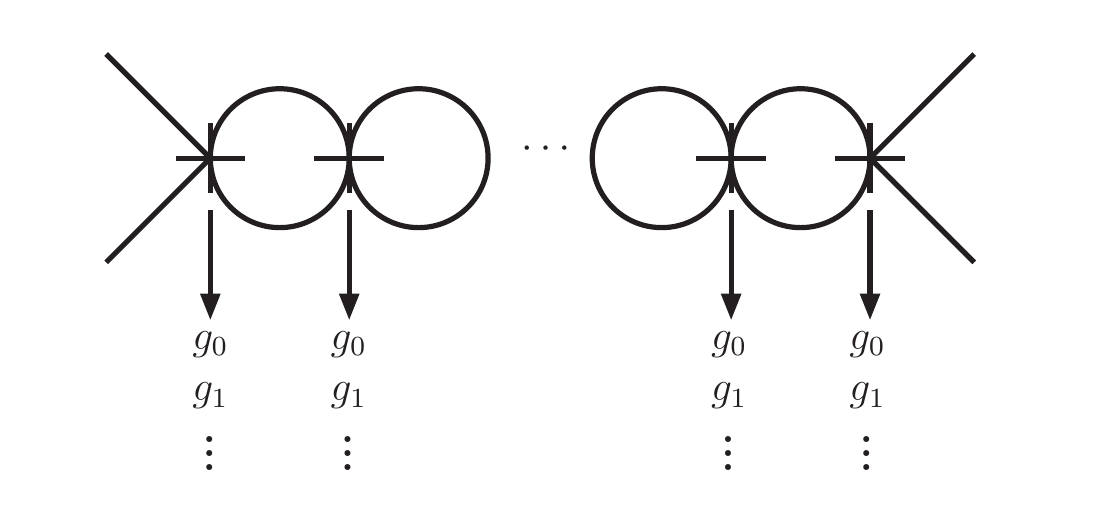}
\caption{Insertions of counterterm Lagrangian vertices for the renormalization of the scattering amplitude}
\label{fig:insertions}
\end{figure}

The dominant contributions of the new terms  to the amplitude  in the $1/N$ counting arise from all possible $g_k$ insertions in diagrams of the form showed in Fig.\ref{fig:insertions}. It is actually not difficult to see that with the covariant structures discussed above, the $\Od(\pif^6)$ terms and higher in \eqref{Lk} give subdominant contributions, which also holds at $T\neq 0$.  As it is explained in  Appendix \ref{app:ren}, each Lagrangian (\ref{Lk}) insertion produces a $s^{k+1}$ power in the vertex at $T=0$. Thus, summing up all the possible $g_k$ insertions in the dominant loop diagrams in Fig.\ref{fig:insertions} is equivalent to the following redefinition of the four-pion vertex \cite{Dobado:1992jg}:

\begin{align}
&\frac{s}{NF^{2}}\rightarrow \frac{s}{NF^{2}}G_{0}(s), \nonumber \\
&G_{0}(s)=\sum_{k=0}^{\infty}{g_k\left(\frac{s}{F^{2}}\right)^{k}},
\label{4piren}
\end{align}
with $g_0=1$, which gives for the $T=0$ amplitude:

\begin{equation}
A(s)=\frac{s}{NF^2}\frac{G_{0}(s)}{1-\frac{s G_{0}(s)}{2F^{2}}J(s)}, 
\label{renormamp0}
\end{equation}
or equivalently, 

\begin{equation}
\frac{1}{A(s)}=\frac{NF^2}{s}\left[\frac{1}{G_0(s)}-\frac{s J(s)}{2F^2}\right],
\label{invamp0}
\end{equation}
written in  a more suitable way to implement its renormalization, as we discuss below.

Now, we can renormalize the bare divergent (and scale independent) LEC $g_{k}$ correctly to absorb order by order the loop divergences coming from the $J(s)$ function.  Thus, we denote   $g^{R}_{k}(\mu)$ for $k\geq 1$ the renormalized (and scale dependent) couplings that are renormalized in terms of the $g_j$ with $j=0,\dots,k$ (see details in Appendix \ref{app:ren}). Equivalently, we define the renormalized function

\begin{equation}
\frac{1}{G_{R}(s;\mu)}=\frac{1}{G_{0}(s)}-\frac{s}{2F^2}J_\epsilon(\mu)
\label{renormG}
\end{equation}
which replaced in \eqref{invamp0} gives rise to the renormalized amplitude:

\begin{equation}
A_{R}(s)=\frac{s}{NF^2}\frac{G_{R}(s;\mu)}{1-\frac{s\,G_{R}(s;\mu)}{32\pi^{2}F^{2}}\ln\left(\frac{\mu^{2}}{-s}\right)}, 
\label{renormamp1}
\end{equation}
where the subscript $R$ is merely added to emphasize that the amplitude is finite. Recall also that the renormalized amplitude is independent of the scale $\mu$, since it was so from the original expression \eqref{renormamp0}, the scale dependence of $J_\epsilon(\mu)$ being cancelled by that of the renormalized $g^{R}_{k}(\mu)$. 

The function $G_{R}(s;\mu)$ can also be written as a formal series in powers of $s$ by expanding both sides of eq.\eqref{renormG} using \eqref{4piren}, so that

\begin{equation}
G_{R}(s;\mu)=\sum_{k=0}^{\infty}{g^R_k(\mu)\left(\frac{s}{F^{2}}\right)^{k}}. 
\label{GRexp}
\end{equation}
Taking $g^{R}_{0}(\mu)=1$ would give the order-by-order renormalization of the $g_{k}$, presented explicitly in Appendix \ref{app:ren} up to $\Od(s^3)$ (specifically \eqref{reng1} and \eqref{reng2}). 

At $T\neq0$, from general grounds, we should be able to renormalize the amplitude with $T=0$ counterterms \cite{lebellac,Kapusta:2006pm}. However, we notice that only with the renormalization of the four-pion vertex in \eqref{4piren} and \eqref{renormG} is not enough to renormalize the thermal amplitude in \eqref{thermalampbare} (even though the divergent part of the $J$ integral is the same as at $T=0$) unless powers of $f(I_{\beta})$ were attached to the counterterm Lagrangians, something that would violate the above mentioned $T\neq0$ renormalization principle. Actually, things become more complicated, since the Feynman rules arising from terms like \eqref{Lk} are not as simple as the $T=0$ ones, which in particular means that a given diagram mixes different $s^{k}$ powers. 

In Appendix \ref{app:ren} we present a detailed analysis of the renormalization scheme that has to be applied to the $T\neq 0$ case. The main conclusion is that the thermal amplitude can be rendered finite with a $T=0$ renormalization where not only the four-point vertex \eqref{4piren} is involved, but also  all $2k+4$-pion vertices of the NLSM Lagrangian \eqref{NLSM} for arbitrary integer $k$ as follows:

\begin{equation}
\frac{s}{(NF^2)^{k+1}}\rightarrow \frac{s}{(NF^2)^{k+1}}G_{0}^{k+1} (s).
\end{equation} 
 
The above renormalization, which can be understood either in terms of an addition of effective diagrams renormalizing the vertices or as a formal renormalization of the metric function at the Lagrangian level (see Appendix \ref{app:ren}), amounts to the redefinition of the effective thermal vertex in Fig.\ref{EffVertx} as given in \eqref{4effren} when the corresponding tadpole diagrams are summed up. In fact, one can also arrive to the same four-pion effective vertex renormalization in \eqref{4effren} starting by redefining the thermal effective vertex with an unknown bare function of $s$ and $T$ that absorbs the divergent part of the $J$ integral in the total amplitude. Thus, we finally have for the thermal amplitude: 

\begin{equation}
A(p;T)=\frac{s G_{0}(s)f[G_{0}(s) I_\beta]}{NF^2}\frac{1}{1-\frac{s G_{0}(s)f[G_{0}(s) I_\beta]}{2F^2}J(p;T)}.
\end{equation}

After using exactly the same renormalization method given in \eqref{renormG}, we obtain a finite renormalized thermal amplitude given by

\begin{equation}
A_R(p;T)=\frac{s G_{R}(s;\mu)f[G_{R}(s;\mu) I_\beta]}{NF^2}\frac{1}{1-\frac{s G_{R}(s;\mu)f[G_{R}(s;\mu) I_\beta]}{2F^2}J_{fin}(p;T;\mu)},
\label{renampT}
\end{equation}
where $J_{fin}$ is the finite part of the thermal $J$ function, defined  in \eqref{JTsepar}. 

Recall that our final finite thermal amplitude \eqref{renampT} is also independent of the $\mu$ scale, as in the $T=0$ case, since the dependence in $J_{fin}$ cancels out that of $G_{R}(s;\mu)$ encoded by the finite renormalization constants $g_{i}^{R}(\mu)$ through \eqref{GRexp}. We point out also that the renormalization scheme discussed here is the same as for $T=0$ and is of course consistent with the previous analysis of the scattering amplitude in the large-$N$ NLSM in \cite{Dobado:1992jg}, but at $T=0$ it is enough to consider the four-point vertex renormalization \eqref{4piren} because the $2k+4$ vertices with $k\geq 1$ simply do not show up in the scattering amplitude  at leading $1/N$ order.

The renormalized thermal amplitude \eqref{renampT} is one of our main results. As it happened in the $T=0$ case \cite{Dobado:1992jg, Dobado:1994fd} the infinite couplings $g_i^R$ parametrize different choices of effective theories sharing basic properties such as renormalizability, the lowest order energy expansion (low-energy theorems) and unitarity (see section \ref{sec:pw}). Indeed, at $T=0$ one can  for instance choose the $g_i^R$ to recover the amplitude of the LSM with explicit exchange of a scalar particle. An alternative approach, which is the one we will follow here, is to fix the scale and the renormalization conditions such as only a finite number of the $g_i^R$ are nonzero. In its simplest version, we can choose $g^R_{k\geq 1}(\mu)=0$. It is not difficult to see that this condition is compatible with the renormalization conditions of the $G_R$ function, namely \eqref{renormG}, and its corresponding Renormalization Group evolution \cite{Dobado:1992jg} and leaves the thermal amplitude  as dependent only on two free parameters, $\mu$ and $F$:

\begin{equation}
A_R(p;T)=\frac{s f[I_\beta]}{NF^2}\frac{1}{1-\frac{s f [I_\beta]}{2F^2}J_{fin}(p;T;\mu)}.
\label{renampTfinal}
\end{equation} 

We will show in section \ref{sec:pwfit} that it is  possible with this approach to fit scattering data fairly well, considering that this is a chiral-limit approach (the finite pion mass case has been analyzed with the same method in \cite{Dobado:1994fd}). That will be enough for our present purposes, since  our main goal is to show that with a $T=0$ amplitude which complies with the above physical requirements, we can obtain a thermal behaviour compatible with different theoretical and lattice expectations regarding chiral symmetry restoration, as we discuss in section \ref{sec:pole}.  

It is also important to stress that following the guide principle that the $T=0$ renormalization should be enough to render a finite amplitude (proven perturbatively up to $\Od(s^3)$ in Appendix \ref{app:ren}), the insertion of counterterms with bare renormalization constants $g_i$, the subsequent absorption of the divergent part of the loop integrals to define the $g^R_i$ and taking  $g_{k>1}^R (\mu)=0$, would have been equivalent to take the thermal amplitude \eqref{thermalampbare} with $J$ replaced by $J_{fin}$. What we have derived here is an explicit construction of such a renormalization scheme.

To end this section and  before proceeding with the analysis of partial waves and thermal poles, we provide a first result related to the pion decay constant at finite temperature. Taking the low-energy limit of the thermal amplitude from its general renormalized form \eqref{renampT} gives simply:

\begin{equation}
A_{R}(p;T)= \frac{s}{NF^{2}}\frac{1}{1-\frac{I_{\beta}}{F^{2}}} + \Od(s^2/F^4),
\end{equation} 
which we can compare with the low-energy expression of the scattering amplitude given by Weinberg's theorem \cite{Weinberg:1978kz} to define a $T$-dependent pion decay constant, namely,

\begin{equation*}
A_{R}(p;T)\equiv\frac{s}{F_{\pi}^{\,2}(T)}+\Od(s^2/F^4),
\end{equation*}
so that:
\begin{equation}
F^2_{\pi}(T)=N F^2\left(1-\frac{T^{2}}{12F^{2}}\right)=F^2_\pi(0)\left[1-\frac{T^{2}}{4F_{\pi}^{\,2}(0)}\right] (N=3).
\label{fpiT}
\end{equation}
where we have used that $F_{\pi}(0)=\sqrt{N}\,F$. The result \eqref{fpiT} coincides with the known ChPT result  to $\Od(T^2)$  \cite{Gasser:1986vb} and with the leading $N$ contribution studied in \cite{Jeon:1996gn}, which are  additional consistency checks of our present analysis. A more careful analysis of $F_\pi$ beyond $\Od(s)$ would require to analyze the residue of the axial-axial correlator \cite{Bochkarev:1995gi,Jeon:1996gn}.

\section{Partial wave analysis and unitarity}
\label{sec:pw}

\subsection{Fitting the $I=J=0$ phase shift at $T=0$}
\label{sec:pwfit}

As discussed above, we will fix the undetermined constants in the scattering amplitude from experimental information. For that purpose and for the subsequent analysis of the $f_0(500)$ pole and chiral restoration, we will consider partial waves with well-defined values of total isospin $I$ and angular momentum $J$, which at $T=0$ are defined in the center of mass (COM) frame $\vec{p}=\vec{0}$:

\begin{equation}
a_{IJ}(s)=\frac{1}{64\pi}\int_{-1}^{1}{T_{I}(s,\cos\theta)}P_{J}(\cos\theta)\,d(\cos\theta),
\label{pwdef}
\end{equation} 
where $P_J$ are Legendre polynomials, $\theta$ is the scattering angle and $T_I$ is a particular combination involving $A(s,t,u)$ (defined in \eqref{generalT}) which gives the scattering amplitude at given isospin $I$, with $t(s,\cos\theta)$, $u(s,\cos\theta)$ in the COM frame. The large-$N$ analysis is specially adequate for the $I=J=0$ channel (see below), which on the other hand is the one we are interested in this work. In that case, we have:

\begin{equation}
T_{I=0}(s,\cos\theta)=NA_{R}(s)+A_{R}(t)+A_{R}(u)=N\left[A_{R}(s)+\Od(1/N)\right],
\label{T0def}
\end{equation}
with $A_R(s)$ given for $T=0$ in \eqref{renormamp1} and where we have made use of the fact that $A(s,t,u)$ depends only on $s$ to leading order in $1/N$ and consistently, we only take the leading order also for the partial wave combination \eqref{T0def}. The final result is  that the $I=J=0$ partial wave  is independent of $N$ (to leading order). At $T=0$ we have then:

\begin{equation}
a_{00}(s)=\frac{s}{32\pi F^2}\frac{G_{R}(s;\mu)}{1-\frac{s\,G_{R}(s;\mu)}{32\pi^{2}F^{2}}\ln\left(\frac{\mu^{2}}{-s}\right)}.
\label{a00T0}
\end{equation}

Recall that within the large-$N$ framework, the other possible isospin channels for pion scattering, namely $I=1$ and $I=2$, are subdominant, since they are proportional to $1/N$. This analysis is  therefore particularly suited for  $a_{00}$  \cite{Dobado:1994fd} which for the thermal case is the most relevant one regarding chiral symmetry restoration from the point of view of  thermodynamic quantities such as the scalar susceptibility, as explained above.

As discussed in section \ref{sec:ren}, we will work within the minimal approach for which $g_k^R(\mu)=0$ for $k\geq 1$ so that we end up with the $T=0$ partial wave in \eqref{a00T0} with $G_R=1$. Defining  the phase shift as customary for elastic channels $a_{IJ}=\vert a_{IJ} \vert e^{i\delta_{IJ}}$, we can use this result to try to fit experimental phase shift data. There are several comments that are pertinent at this point: first, the choice of data sets is delicate because there have been several experiments over the years with results sometimes incompatible among them for this channel.  In addition, we have to take into account that we are working in the chiral limit, so that we expect our amplitude to describe more naturally data sufficiently away from threshold, i.e., typically for  large $\sqrt{s}/(2m_\pi)$. On the other hand, there is also a natural upper limit of applicability for $\sqrt{s}$, namely below the next resonance mass in this channel, which is the $f_0(980)$. This would need  the inclusion of the strange sector. Thus, we have chosen as data for the fit the sets given by \cite{grayer} in the $\sqrt{s}$ range 450-800 MeV. In addition, we consider also the parametrization of the scattering amplitude described in \cite{pelaezprl} in the same energy region. That parametrization provides a precise description of $\pi\pi$ scattering from a combined analysis based on dispersion relations and provides an accurate prediction of the $f_0(500)$ and $f_0(980)$ pole parameters. We do not include in the fit the  recent (and also more precise) data of the NA48 experiment \cite{NA48} which are  very low-energy data below $\sqrt{s}\sim$ 400 MeV. Those low-energy data are very well described  by the  parametrization \cite{pelaezprl}. In the fit to the parametrization \cite{pelaezprl}, we select points with a 5 MeV energy interval and take into account the small  uncertainties given in  that paper.

Another important point is that in principle the values obtained for $F$ in our fits should not be far from the physical value of the pion decay constant $F_\pi=\sqrt{N}F$. In the chiral limit, $F_\pi\simeq$  87 MeV \cite{Gasser:1983yg,Bijnens:2011tb}, so that $F$ would be around $F_\pi/\sqrt{3}\simeq$ 51 MeV. However, once again it is important to stress that we are forcing our chiral limit amplitude to fit data for massive pions and it is then not surprising that we need a  higher value for $F$ since mass corrections increase the $F_\pi$ value, so that we are encoding in $F$ a great part of the uncertainty we have in our chiral limit analysis. 

In any case, it is  not the purpose of this work to provide a very precise fit to experimental data, as  in other unitarized or dispersive approaches \cite{pelaezprl,GomezNicola:2001as} which can even be compared to lattice results by suitable mass extrapolations \cite{Nebreda:2011di}. After all, this is just a two-parameter fit in the chiral limit. We just need some reasonable reference values for the parameters such as we generate dynamically a pole in the second Riemann sheet with consistent values for the pole position, so that the thermal behaviour corresponds to a physically realistic situation. 

\begin{figure}
\centering
\includegraphics[scale=0.4]{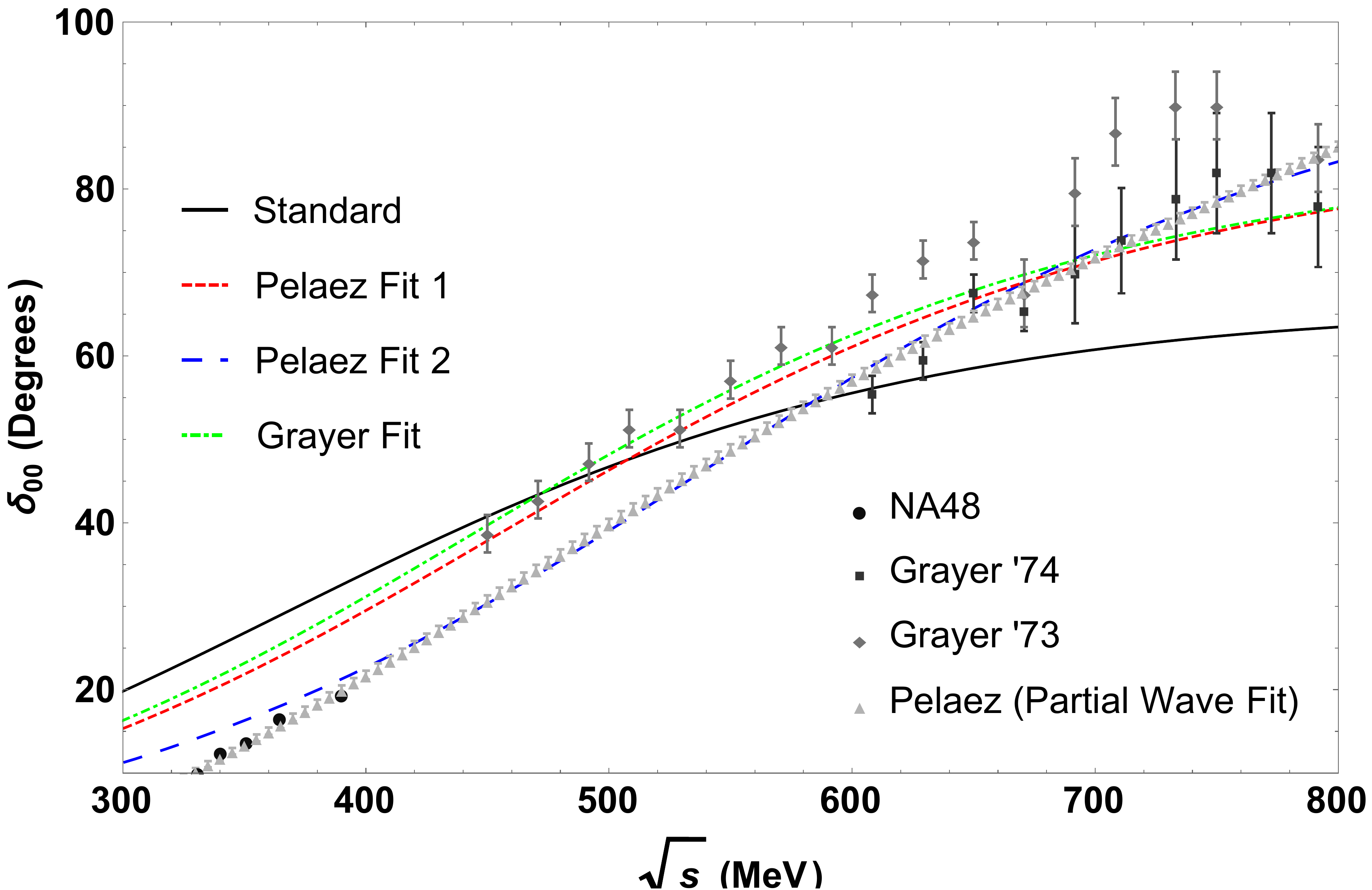}
\caption{$I=J=0$ channel phase shift for  different fits, as explained in main text. Pel\'aez parametrization is given in \cite{pelaezprl}, NA48 low-energy data in \cite{NA48} and Grayer data in \cite{grayer}.}
\label{phaseshift}
\end{figure}

In Fig.\ref{phaseshift} we show the $I=J=0$ phase shift as a function of the COM energy obtained with our large-$N$ amplitude with our best fits to Grayer data and to Pel\'aez et al parametrization.  The corresponding fit parameters are given in Table \ref{table:fittable}.

\begin{table}[h]
\begin{tabular}{cccc} \hline \hline
	Parameters & Grayer & Pel\'aez 1 & Pel\'aez 2 \\ \hline
	$F\pm\Delta F\text{ (MeV)}$  & 63.16$\pm$1.62 & 65 (fixed) & 75.98$\pm$ 0.16\\ 
	$\mu\pm\Delta\mu\text{ (MeV)}$ & 1523.35$\pm$143.34 & 1607.89$\pm$3.62& 2763.51 $\pm$ 23.81 \\ 
	$R^{2}$ & 0.9958 & 0.9951 & 0.9999  \\ \hline \hline
\end{tabular}
\caption{Parameters for the Grayer and Pel\'aez  data fits  and their respective coefficients of determination.}
\label{table:fittable}
\end{table}

The behaviour in this region is typically flat, compatible with having a wide resonance far from the real axis, in contrast with the $I=J=1$ channel where the presence of the $\rho(770)$ narrow resonance is clearly evident for real $s$ \cite{GomezNicola:2001as}.  The fits are less sensitive to the value of $\mu$, which is natural taking into account that the dependence with that parameter is logarithmic. As commented above, the values for $F$ are rather high, compared with the expected  value, which  is a consequence of dealing with the chiral limit. In the case of the fit to the points in \cite{pelaezprl}, this effect is particularly notorious in the fit named ``Pel\'aez 2". However, we present the results of that fit anyway, because of its remarkably good accuracy to reproduce the parametrization \cite{pelaezprl},  even in the very-low energy region, where as commented the approach was not meant to be applicable. In that sense, this is the fit that describes better the scattering data for this channel, although the price to pay is an unnatural deviation in the pion decay constant. The chiral limit restriction, as well as other possible effects suppressed in the large-$N$ limit, such as the $t,u$ dependence of the amplitude, are encoded in that $F$ value.  Alternatively, in the fit named ``Pel\'aez 1", we fix the value $F$=65 MeV such that we get a similar fit quality as the Grayer one, parametrized by $R^2$.  Consequently, the values of $\mu$ obtained in those fits also remain  close to each other. The uncertainties  in $F$ and $\mu$ given in Table \ref{table:fittable} only include the error in the fit to the selected data, and are therefore clearly underestimates, taking into account the additional sources of uncertainty mentioned above.  In this context, it is also useful to compare with the values obtained in \cite{Dobado:1994fd} for a fit including mass corrections, giving $F=\text{55.41 MeV and }\mu=775\text{ MeV}$. We will denote the latter values as ``standard values", whose corresponding curve with our chiral-limit amplitude is also depicted in Fig.\ref{phaseshift}. It is not surprising that this curve does not fit the data properly, because the parameters are taken from the massive case fit. Finally,  to check the robustness of our approach we have also tried to fit the same data sets by including one nonzero additional parameter $g_1^R$ in $G_R$ so that we have now a three-parameter fit. The result is that the best fit yields values for $F$ and $\mu$ very close to those in Table \ref{table:fittable} with $g_1^R$ of the expected natural order for the LEC \cite{Bijnens:2011tb} but compatible with zero. This is a consistency check of this approach, reinforcing also the idea commented in the introduction that the $I=J=0$ channel is less sensitive to the LEC than to the loop effects.

\subsection{Unitarity at zero and finite temperature and the $f_0(500)$ pole}
\label{sec:unit}

At $T=0$ the unitarity condition for partial waves  in elastic pion scattering reads $\im a_{IJ}(s+i0^+)=\sigma(s,m_\pi)\vert a_{IJ}(s) \vert^2$ for $s\geq 4m_\pi^2$ (two-pion threshold), where $\sigma$ is the two-pion phase space:

\begin{equation}
\sigma(s,m_\pi)=\sqrt{1-\frac{4m_\pi^2}{s}}.
\end{equation}

Equivalently, $T=0$ unitarity reads $\im \left[a_{IJ}^{-1}(s+i0^+)\right]=-\sigma(s,m_\pi)$.  It is not difficult to see that the large-$N$  $a_{00}$  partial wave at $T=0$ given in \eqref{a00T0} satisfies this condition (recall that $G_R(s)$ is a real function) in the chiral limit:

\begin{equation}
\im\left[\frac{1}{a_{00}(s+i 0^+)}\right]=-\frac{1}{\pi}\im\left[\ln\left(\frac{\mu^2}{-s-i 0^+}\right)\right]=-1=-\sigma(s,0) \quad \mbox{for} \quad s\geq 0.
\end{equation}

Exact unitarity is one of the prominent features of the large-$N$ approach. Recall that in the standard ChPT series, unitarity holds only perturbatively order by order and demanding exact unitarity is what leads for instance to the IAM method. 

What is even more interesting for our purposes  is that there is a thermal unitarity relation which holds perturbatively in ChPT \cite{GomezNicola:2002tn}, and is given by $\im a_{IJ}(s+i0^+;T)=\sigma_T(s,m_\pi)\vert a_{IJ}(s;T) \vert^2$ for $s\geq 4m_\pi^2$, where the partial waves at finite temperature are defined in the center of momentum frame $\vec{p}=\vec{0}$, i.e., the frame in which pions are at rest with the thermal bath, and where the thermal phase space $\sigma_T$ is:

\begin{equation}
\sigma_T(s,m_\pi)=\sigma(s,m_\pi)\left[1+2n_B\left(\frac{\sqrt{s}}{2}\right)\right].
\label{thermalphsp}
\end{equation}

The Bose-Einstein correction in \eqref{thermalphsp} can be interpreted as the difference of enhancement and absorption of scattering states in the thermal bath \cite{GomezNicola:2002tn} . If this thermal perturbative relation is imposed to hold also for the full amplitude, one ends up with a unitarized thermal amplitude which gives rise to the $T$-dependence of the $f_0(500)$ and $\rho(770)$ thermal poles   \cite{Dobado:2002xf,FernandezFraile:2009mi,Nicola:2013vma}.  

An important result of the present work is that the thermal unitarity relation holds exactly for the  large-$N$ scattering amplitude, thus providing theoretical support to the previously mentioned works on this subject. This can be readily checked from our previous results. From the definition of partial waves in \eqref{pwdef} and \eqref{T0def}, now with the thermal amplitude $A_R$
 in \eqref{renampT} at $\vec{p}=\vec{0}$, i.e., with $J_{fin}$ given by \eqref{Jthermal1}-\eqref{Jthermal2}, we have:

 \begin{equation}
a_{00}(s;T)=\frac{s G_{R}(s;\mu)f[G_{R}(s;\mu) I_\beta]}{32\pi F^{2}}\frac{1}{1-\frac{s G_{R}(s;\mu)f[G_{R}(s;\mu) I_\beta]}{32\pi^2F^2}\left[\ln\left(\frac{\mu^{2}}{-s}\right)+16\pi^2\delta J(s;T)\right]}.
\label{a00T}
\end{equation}

Using  \eqref{imJther}, we get now:

\begin{equation*}
\text{Im}\left[\frac{1}{a_{00}(s+i0^+;T)}\right]=-\frac{1}{\pi}\left[\pi+16\pi^{2}\text{Im}\delta J(s;T)\right]=-\left[1+2n\left(\frac{\sqrt{s}}{2}\right)\right]=-\sigma_T(s,0),
\end{equation*}
which is the thermal unitarity relation.

Unitarity allows to define the Riemann second-sheet partial wave, both at $T=0$ and at $T\neq 0$,  when the amplitude is continued analytically to the $s$ complex plane  so that $\im a^{II}(s-i0^+)=\im a (s+i0^+)$ for $s>4m_\pi^2$. This is achieved by $a^{II}(s;T)=a(s;T)/\left[1-2i\sigma_T a(s;T)\right]$. The second-sheet amplitude presents poles which correspond to the physical resonances, which in the case of pion scattering are the $f_0(500)$ ($I=J=0$) and $\rho(770)$ ($I=J=1$). The $T$-dependent poles can be extracted numerically by searching for zeros of $1/a^{II}(s;T)$ in the $s$ complex plane. We denote the pole position as customary by $s_p(T)=\left[M_p(T)-i\Gamma_p(T)/2\right]^2$. 

\begin{figure}
\begin{center}
 \centerline{\includegraphics[width=.4\linewidth]{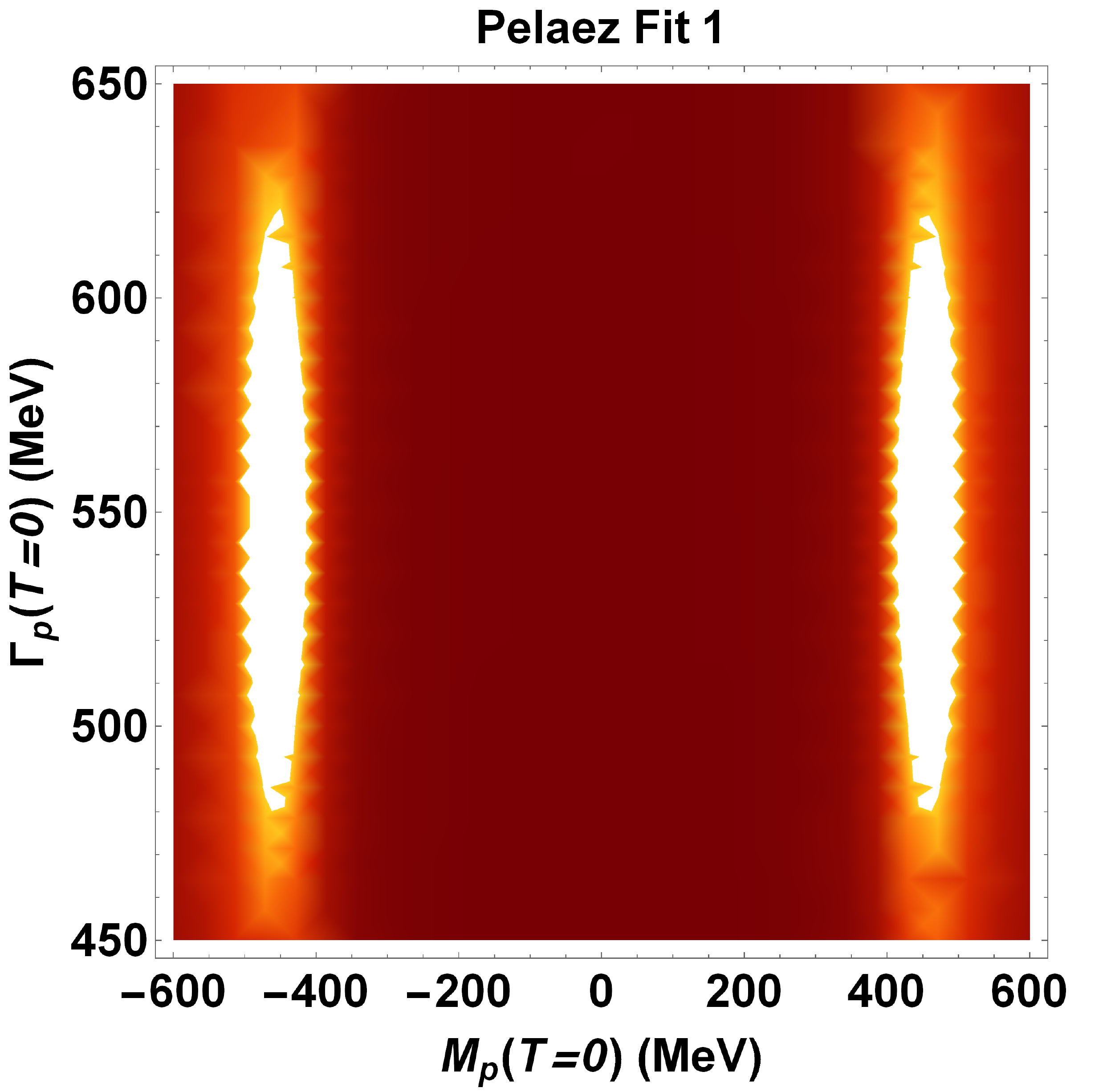} \includegraphics[width=.4\linewidth]{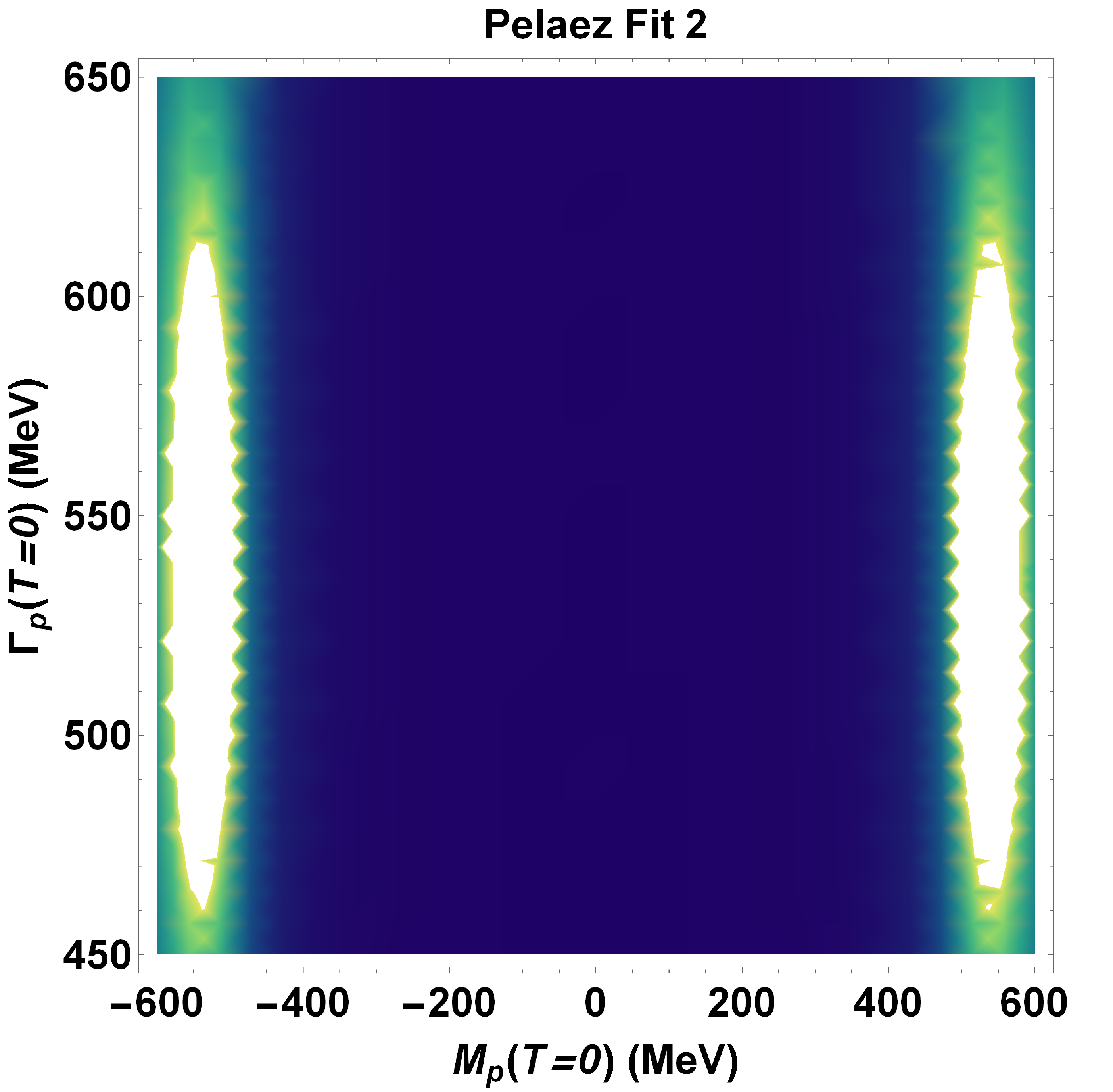}}
 \centerline{\includegraphics[width=.4\linewidth]{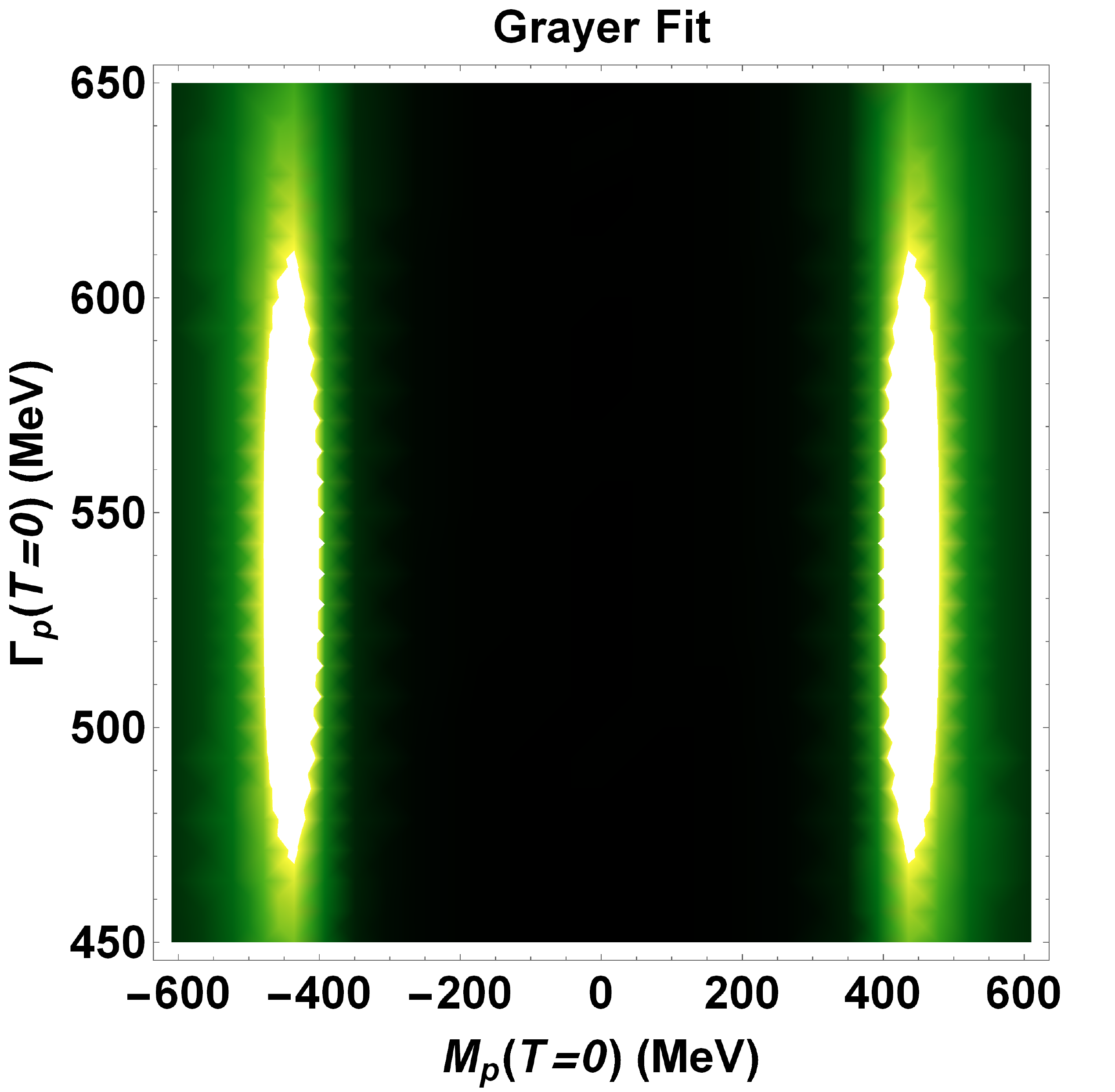} \includegraphics[width=.4\linewidth]{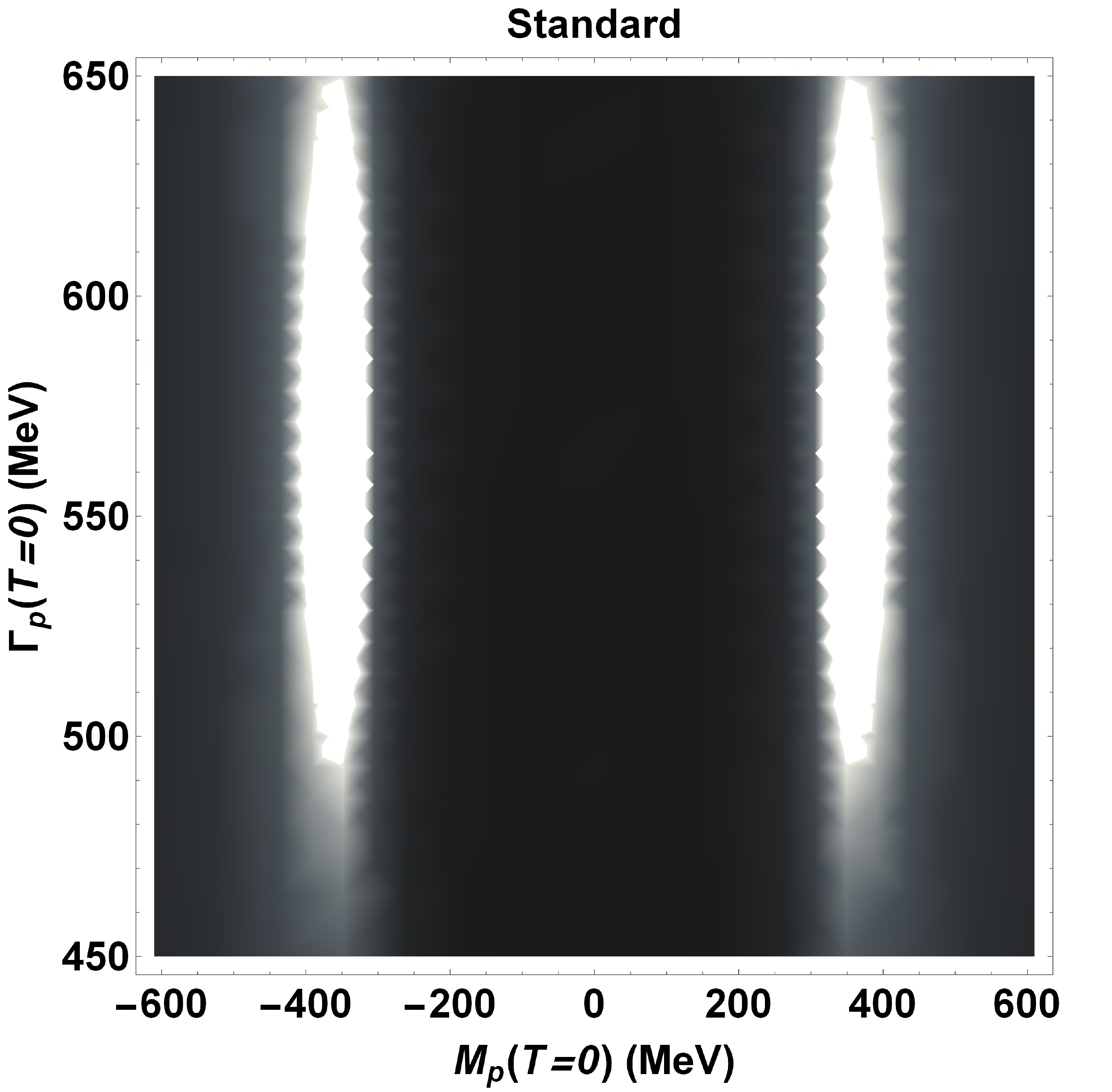}}
\end{center}
\caption{Surface levels for $|a_{00}^{\,II}(s;T)|^{2}$ at $T=0$ with Pel\'aez 1,2, Grayer and Standard fit parameters respectively. The elliptic regions show the positions of the pole in the upper half of the second Riemann sheet.}
\label{fig:surfacelev}
\end{figure}

In the next section, we will give the detailed results of the thermal pole evolution within our present large-$N$ approach. Before that, in Table \ref{table:poletable} we give the values of the $T=0$ $f_0(500)$ pole, from the partial wave in \eqref{a00T0} taking $G_R=1$,  with the different parameter sets of Table \ref{table:fittable} and in Fig.\ref{fig:surfacelev} we provide the surface-level plots for those poles.  For comparison, we also present the results of the IAM in the chiral limit, using the same LEC as in previous works \cite{Nicola:2013vma}.

\begin{table}
\begin{tabular}{ccc}   \hline \hline
	Fit & $M_{P}(T=0)$ & $\Gamma_{P}(T=0)$  \\ \hline 
	Grayer  & 438.81 & 536.47 \\ 
	Pel\'aez 1& 452.42 & 546.26 \\
	Pel\'aez 2& 535.53 & 534.59 \\
	IAM & 406.20 & 522.70 \\
	Standard & 356.97 & 566.05 \\ \hline \hline
	\end{tabular}
\caption{Values for masses and widths (in MeV) of the $f_{0}(500)$ pole at zero temperature.}
\label{table:poletable}
\end{table}

These values can be compared for instance with those obtained  in the analysis \cite{pelaezprl} and given by $M_p=457^{+14}_{-13}$ MeV, $\Gamma_p=558^{+22}_{-14}$ MeV, compatible also with the PDG values $M_p\simeq$ 400-500 MeV, $\Gamma_p=400-700$ MeV \cite{Agashe:2014kda}, with  a large uncertainty and where the results of different analysis can be found. We refer also to the recent review \cite{Pelaez:2015qba} for updated results on the $f_0(500)$ pole parameters. 
The values we obtain here are compatible with those typically  quoted in the literature, which is remarkable given the uncertainties explained above related mostly to our chiral limit description. This is an important step in our analysis since we want our $T=0$ values for the pole to be as close as possible to realistic values, so that we can track its temperature evolution trustfully.  In fact, we see that having already paid the price of increasing somewhat the value of $F$, the values for the mass and pole position are not far from those expected in the physical case. As a rule of thumb, we would expect that in the chiral limit, the pole mass would decrease (as it does the quark-like component of this state) and the pole width would increase by a phase-space argument. That is the case for instance for the IAM pole, which in the massive case is at  $M_p=441.47$ MeV, $\Gamma_p=464.34$ MeV  with the same LEC that give rise to the massless pole quoted in Table \ref{table:poletable}. This is also the reason why the results in Table  \ref{table:poletable} for the ``Standard" values, which correspond to a massive-pion fit, give in the chiral limit a smaller mass and higher width than the other large-$N$ fits.

\section{Thermal evolution of the pole and connection with chiral symmetry restoration}
\label{sec:pole}

From our previous discussion, we can now follow the temperature evolution of the $f_0(500)$ pole and compare with previous analysis. In addition, following the proposal in \cite{Nicola:2013vma}, the thermal pole can be connected with chiral symmetry restoration (in the chiral limit) via the scalar susceptibility. 

\begin{figure}
\centering
\includegraphics[scale=0.35]{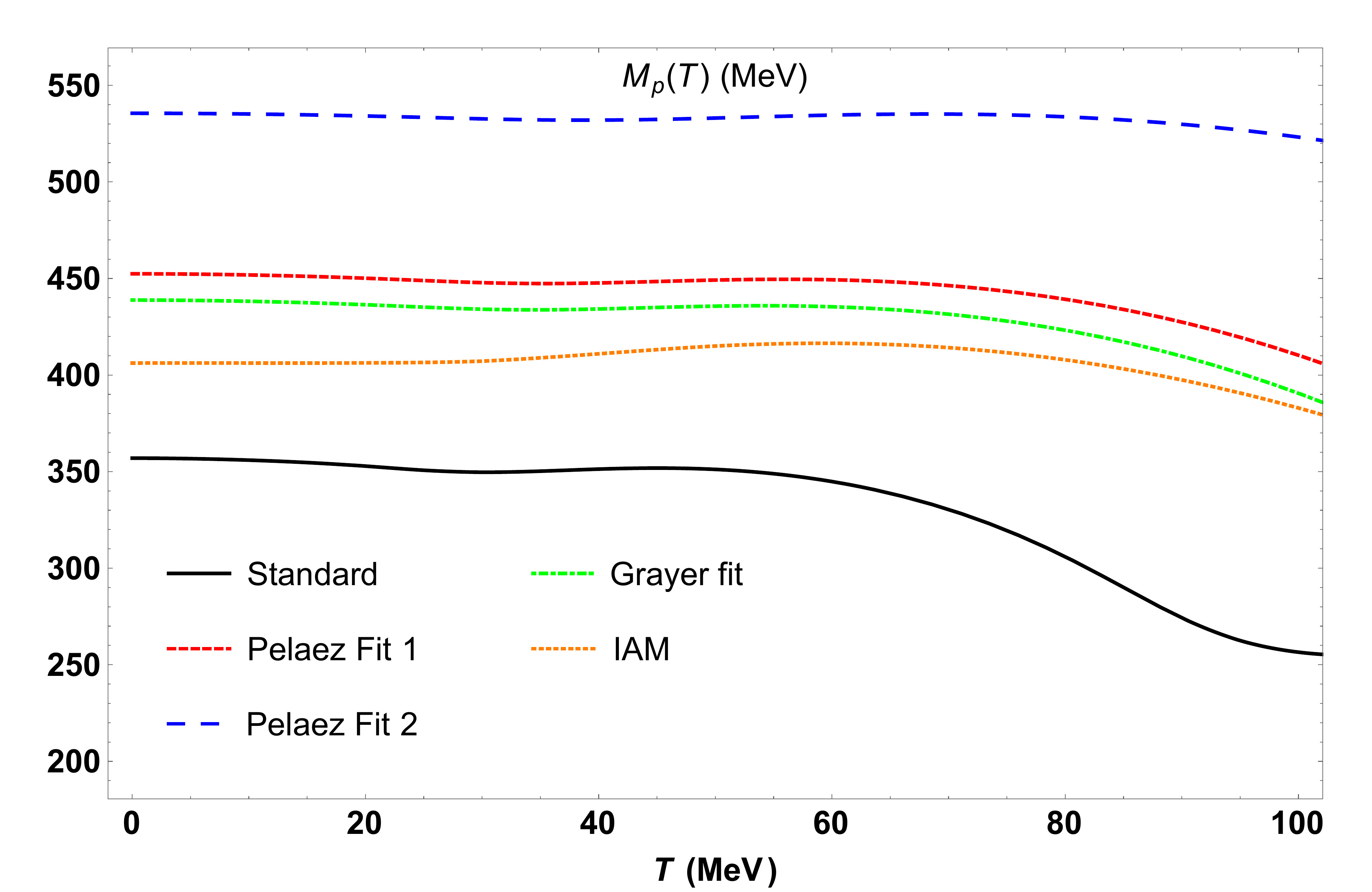}
\caption{Mass of the $f_{0}(500)$ pole as functions of temperature when considering different fits in the large-$N$ framework. We also compare with  the IAM approach in the chiral limit.}
\label{fig:masspole}
\end{figure}

\begin{figure}
\centering
\includegraphics[scale=0.35]{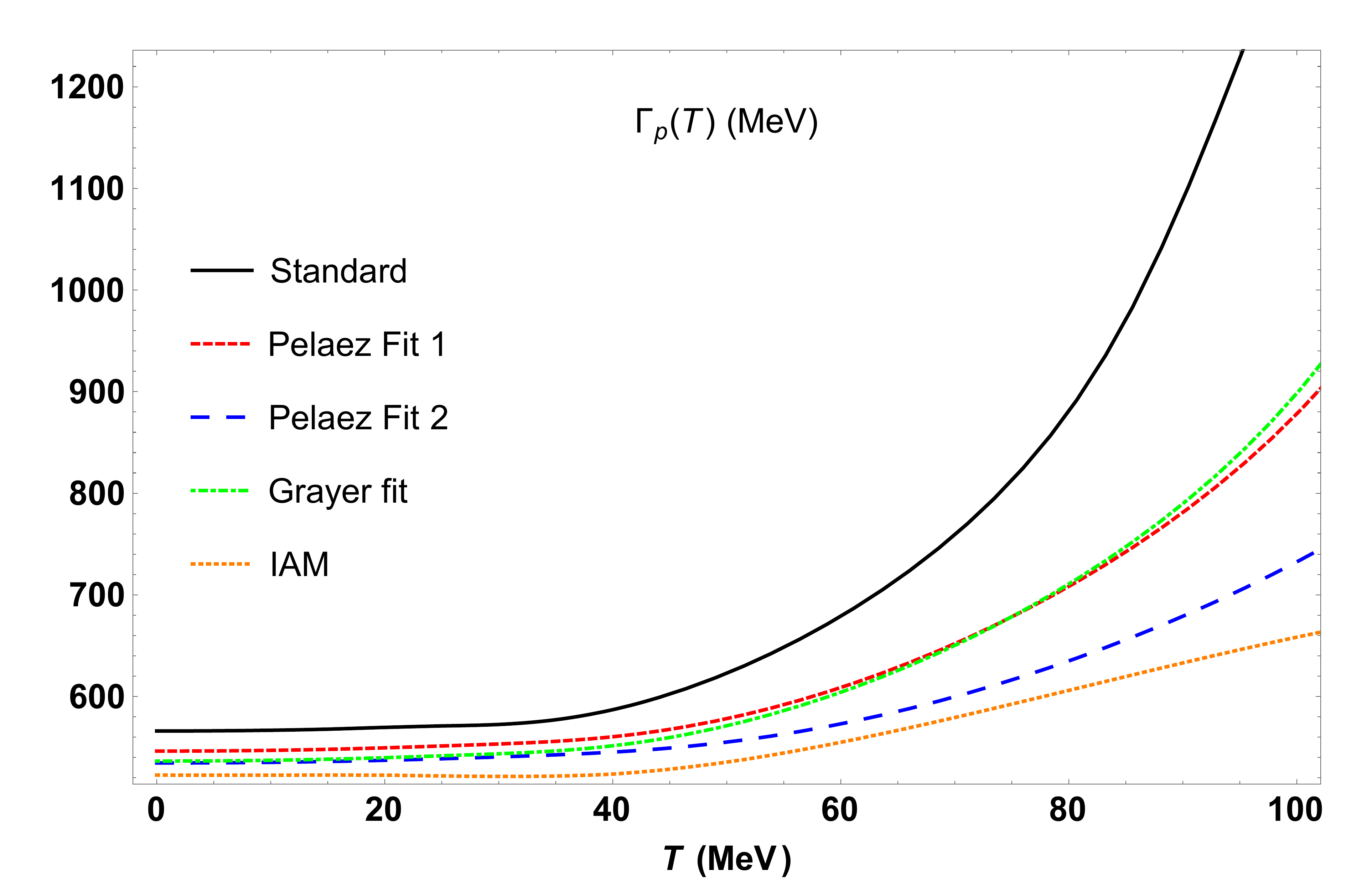}
\caption{Width of the $f_{0}(500)$ pole as functions of temperature when considering different fits and  the IAM  in the chiral limit.}
\label{fig:widthpole}
\end{figure}

The results we obtain for the pole position parameters $M_p(T)$, $\Gamma_p(T)$ in the second Riemann sheet at finite $T$ from the thermal partial wave in \eqref{a00T} (with $G_R=1$) are given in Figs. \ref{fig:masspole} and \ref{fig:widthpole}. We also compare with the IAM approach in the chiral limit. A general tendency is observed regardless of the approach and the parameters, and is that the pole mass decreases with $T$ while the pole width increases. Thus, in the chiral limit at finite temperature, the $f_0(500)$ remains a wide resonance below the chiral transition. However, there are significant quantitative deviations when comparing different parameter sets, the results with the ``Pel\'aez 1" and ``Grayer" fits and the IAM remaining reasonably close together.


What is more revealing is the behaviour of $M_S^2=M_p^2-\Gamma_p^2/4$. This is nothing but the real part of the  self-energy of the effective scalar state exchanged in pion scattering. On the other hand, the scalar susceptibility $\chi_S(T)=-\partial\langle \bar q q \rangle/\partial m_q$, with  $\langle \bar q q \rangle$ the quark condensate and $m_q$ the quark mass, is defined as the zero four-momentum scalar correlator and is  saturated precisely by $M_S^2$, assuming that the real part of the self-energy does not vary much in momentum from $p^2=0$ to $p^2=s_p$ \cite{Nicola:2013vma}. This is specially relevant close to the critical region where $M_S^2$ is expected to vanish, so that:

\begin{equation}
\frac{\chi_S(T)}{\chi_S(0)}=\frac{M_S^2(0)}{M_S^2(T)}=\frac{M_p^2(0)-\Gamma_p^2(0)/4}{M_p^2(T)-\Gamma_p^2(T)/4}.
\label{scalarsus}
\end{equation}

Moreover, in \cite{Nicola:2013vma} it has been shown that using the IAM scalar $f_0(500)$ thermal pole to saturate the scalar susceptibility through \eqref{scalarsus} generates precisely a unitarized version of $\chi_S$ which develops a maximum very close to the critical point predicted by lattice analysis, i.e. $T_c\sim$ 155 MeV, in the physical massive case where the transition is believed to be a crossover. In that case, the maximum comes from a combination of the dropping $M_p(T)$ behaviour  and the $\Gamma_p(T)$ behaviour, which grows at low and moderate temperatures due to phase space increasing but drops near the transition where mass reduction is dominant. In our present approach in the chiral limit, $\Gamma_p(T)$ grows monotonically, so that the thermal phase space dominates over mass reduction in  the temperature range of interest. 

\begin{figure}
\centering
\includegraphics[scale=0.35]{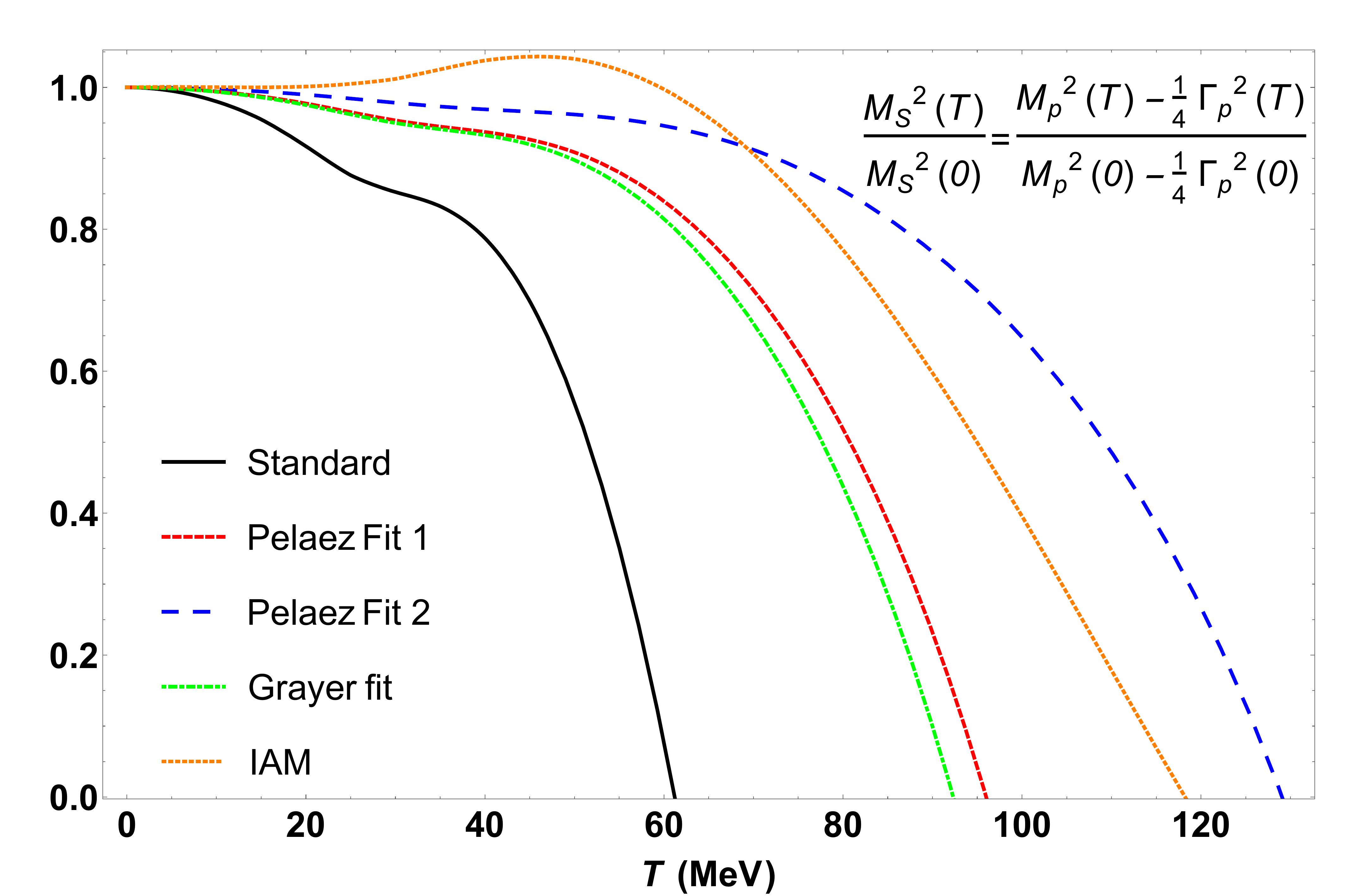}
\caption{Normalized scalar mass squared  (inverse scalar susceptibility) as function of the temperature for different parameter sets and the IAM.}
\label{scalarsuscep}
\end{figure}

 In the chiral limit the transition should be a second-order one, so that such maximum should become a pole, accompanied with a significant reduction in $T_c$. We show our results for $M_S^2(T)/M_S^2(0)$ in Fig.\ref{scalarsuscep}. A clear dropping behaviour vanishing at $T_c$ is observed, corresponding  to a chiral restoration second-order continuous phase transition, according to our previous discussion.  The values of $T_c$ obtained for different parameters are given in Table \ref{temptable}. We also compare with the IAM in the chiral limit, which shows a similar dropping behaviour, although qualitatively different in the intermediate temperature region. 

\begin{table}
\begin{tabular}{ccc}   \hline \hline
	Parameter set & & $T_c$  (MeV) \\ \hline
	Grayer  & & 92.33   \\ 
	Pel\'aez 1  & & 96.00 \\ 
	Pel\'aez 2  & & 129.07 \\
	IAM & & 118.23\\
	Standard & & 61.20 \\ \hline \hline
\end{tabular}
\caption{Values for the chiral critical temperature obtained for  different parameter sets and the IAM.}
\label{temptable}
\end{table}

Let us comment now on these results. A first interesting consistency check, from the formal point of view, is that the result for $T_c$ we obtain here is independent of $N$ for large $N$, since it is extracted from a $N$-independent quantity, namely the  partial wave \eqref{a00T}. This is consistent with the $T_c$ extracted from the partition function, which to leading order in $1/N$ is also $N$-independent and is given by $T_c^2=12F^2$ in the chiral limit \cite{Bochkarev:1995gi,Andersen:2004ae}.  That happens also in other approaches such as ChPT, where the thermal loop corrections to the quark condensate increase proportionally to $N$ but are divided by $F_\pi^2\sim NF^2$ \cite{Gasser:1986vb,Gerber:1988tt}. However, our numerical values for $T_c$ extracted in the way we have just discussed are remarkably closer to the range expected from lattice simulations than the large-$N$ value just mentioned. As commented in the introduction, phenomenologically we expect a $T_c$ value of about 80\% of the massive case, namely  around 120 MeV. In addition, the predictions from our large-$N$ approach are very close to the IAM one, which is formulated  for $N=3$. Thus, with our approach we obtain results closer to the real $N=3$ world, even though they come from the leading order in $1/N$. Generically speaking, we would expect up to 30\% uncertainties for $N=3$, coming from the neglected $1/N$ corrections, but we see that our results are even better than this.  The key point to understand this is that, apart from the large-$N$ resummation, which incorporates important formal properties such as thermal unitarity, we have chosen our parameters to obtain reliable values for the phase shifts and pole at $T=0$, i.e., close to the physical case. In this sense, it is important  to remark  that getting $T=0$ pole values quite close to the physical (massive) ones, by increasing the $F$ value, does not imply  that the $T$-evolution of the pole towards chiral restoration  should be like the massive case, e.g. for  $T_c$, since there are genuine  massive thermal effects that we are neglecting when taking the chiral limit, like the combined dependence of thermal distribution functions on mass and temperature \cite{Gerber:1988tt}.  For that reason, we get $T_c$ values closer to the expected chiral limit ones. We also mention at this point that studies of the chiral phase transition based on Renormalization Group  yield $T_c\simeq$ 100.7 MeV in the chiral limit \cite{Berges:2000ew}, also very close to our present analysis.

\begin{table}
\begin{tabular}{ccccc}   \hline \hline
	Fit & & $\gamma_\chi$ & & $R^{2}$ \\ \hline
	Grayer & & 0.875 & & 0.99987 \\ 
	Pel\'aez 1 & & 0.938 & & 0.99997 \\
	Pel\'aez 2 & & 0.919 & & 0.99995 \\
	IAM & & 1.012 & & 1\\
	Standard & & 0.842 & & 0.99728 \\ \hline \hline
	\end{tabular}
\caption{Critical exponents for $\chi_S$ extracted from our results in Fig.\ref{scalarsuscep}}
\label{tab:scaling}
\end{table}

Another chiral-restoration property we can examine is the scaling law for the scalar susceptibility defined through \eqref{scalarsus}, i.e., calculate the critical exponent $\gamma$ determined as $\chi_S(T)/\chi_S(0)\sim (T_c-T)^{-\gamma_\chi}$ for $T\rightarrow T_c^-$. The results for the best fits are showed in Table \ref{tab:scaling}.  We can compare this analysis, on the one hand, with the exact result for the nonlinear $O(N)$ model for $N\rightarrow\infty$ in four dimensions, $\gamma_\chi^{O(\infty)4D}=1+\Od(1/N^2)$ \cite{zjbook}. On the other hand, the critical exponent of the  $O(4)$ three-dimensional Heisenberg model,  which lattice QCD results resemble within uncertainties \cite{Ejiri:2009ac}, is $\gamma_\chi^{O(4)3D}\simeq $ 0.54 for $T<T_c$ in the chiral limit \cite{Pelissetto:2000ek,Ejiri:2009ac}. Our results and the IAM one lie close to those values,  providing then a consistency check of our approach to define the scalar susceptibility from the $f_0(500)$ pole. 

\section{Conclusions}
\label{sec:conc}

We have studied pion scattering in the large-$N$ $O(N+1)/O(N)$ model at finite temperature in the chiral limit and its consequences regarding the $f_0(500)$ pole and chiral symmetry restoration. Our analysis gives rise to interesting theoretical and phenomenological results, consistent with previous analysis and lattice data.

After calculating the relevant Feynman diagrams, which include an effective thermal vertex from tadpole resummation, an important part of our work has been devoted to show that it is possible to find a renormalization scheme rendering the thermal amplitude finite with a $T=0$  renormalization of the corresponding vertices. This is a nontrivial extension of the $T=0$ renormalization of the scattering amplitude, since the breaking of Lorentz covariance in the thermal bath induces  crossed terms between  tadpole-like and  $J_T$ loop functions. In the low-energy expansion of the model, up to $\Od(s^3)$, we have checked explicitly  this renormalization scheme, providing a diagrammatic and Lagrangian interpretation. 

Another relevant result is that the large-$N$ thermal amplitude satisfies exactly the thermal unitarity relation, imposed in previous works as a physical condition for the exact amplitude.  Its low-energy properties are also preserved, being consistent for instance with the thermal dependence of the pion decay constant.

By a suitable choice of the low-energy constants, similarly to the $T=0$ case, compatible with the scale evolution of the renormalized couplings, we end up with a phenomenological unitary amplitude depending only on two parameters, $F$ and $\mu$. By fitting those parameters to experimental data in the $I=J=0$ channel, which is more reliable for data not very close to threshold in the elastic region, we reproduce the pole position of the $f_0(500)$ in the second Riemann sheet fairly consistently with PDG values and recent determinations. The chiral limit character of our approach implies a larger value for $F$ than phenomenologically expected, but it allows to obtain pole position parameters $M_p,\Gamma_p$ closer to the physical case. The fits to data are actually very good in the chosen region, precisely the most relevant energy range concerning this resonant state. 

Once the $T=0$ pole has been fixed to physical values, we have studied its evolution with temperature. The $f_0(500)$ pole remains a wide state for all the temperature range of interest, the real and imaginary parts $M_p(T)$ and $\Gamma_p(T)$ behaving similarly to the IAM analysis, showing the signature of chiral restoration.  In order to explore this further,  we define a scalar susceptibility $\chi_S(T)$ saturated by the inverse of $M_S^2(T)=M_p^2(T)-\Gamma_p^2(T)/4$, corresponding to the real part of the scalar state self-energy at zero four-momentum, which diverges at a given $T_c$ with a power law, as it corresponds to a continuous second-order phase transition in the chiral limit. The values obtained for $T_c$, as well as the critical exponent of $\chi_S$, are consistent with those obtained with other analytical approaches, such as the IAM, and with lattice analysis, being compatible with a $O(4)$  scaling. The combination of the large-$N$ framework with the phenomenological features of the $f_0(500)$ pole allows to improve the predictions of previous  approaches based on the partition function. Thus, we obtain a very reasonable description of the chiral restoration transition within this approach, given the different uncertainties involved, such as possible $1/N$ corrections near the physical $N=3$ case or the absence of heavier degrees of freedom, which should play an important role near the transition and improve our simple pion gas scenario.

\section*{Acknowledgments}
Work partially supported by the Spanish Research contracts FPA2011-27853-C02-02, FPA2014-53375-C2-2-P. We also acknowledge the support
of the EU FP7 HadronPhysics3 project, the Spanish Hadron Excelence Network (FIS2014-57026-REDT) and the UCM-Santander project GR3/14 910309. Santiago Cort\'es thanks Prof. Jos\'e Rolando Rold\'an and the High Energy Physics group of Universidad de los Andes and COLCIENCIAS for financial support. We are also grateful to Jos\'e Ram\'on Pel\'aez and Jacobo Ruiz de Elvira for useful comments and for providing us with their parametrization values for the phase shift. 

\appendix

\section{Details of the renormalization procedure}
\label{app:ren}

A Lagrangian of the form (\ref{Lk}) gives rise to the Feynman rule $2^k\left[(p_A\cdot p_B)^k (p_C\cdot p_D)+(p_A\cdot p_B) (p_C\cdot p_D)^k\right]\delta_{AB}\delta_{CD}$ where $p_{A,B,C,D}$ are the four-momenta of the four legs and $A,B,C,D$ their isospin indices. We will consider insertions of these counterterms in diagrams of the form depicted in Fig.\ref{fig:insertions}, which will be the dominant ones in $1/N$.  For those insertions, we will have to deal then with integrals of the type:

\begin{equation}
J_n(p,T)=\int_T d^D q\frac{\left[q\cdot(p-q)\right]^n}{q^2(p-q)^2}=\frac{1}{2^n}\int_T d^Dq \frac{\left[s-q^2-(p-q)^2\right]^n}{q^2(p-q)^2} \qquad (n=0,1,2\dots),
\label{Jn}
\end{equation}
where $\int_T d^Dq$ is short for  $\displaystyle T\sum_n \int\frac{d^{D-1} \vec{q}}{(2\pi)^{D-1}}$, $q_0=i\omega_n$, $p_0=i\omega_m$ and $s=p^2$ (after analytic continuation). The case $n=0$ corresponds to $J_0=J(i\omega_m,\modp;T)$ in \eqref{JT}.

First, consider the $T=0$ case. Since $\int dq q^\alpha=0$ in DR \cite{Leibbrandt:1975dj}, the only remaining terms after expanding the numerator in (\ref{Jn}) are the $s^n$ one and the contributions  $\int\frac{q^{2j}}{(p-q)^2}=\int\frac{(p+q)^{2j}}{q^2}$ and $\int\frac{(p-q)^{2j}}{q^2}$. The latter vanish also in DR since $\int dq \frac{q_1^{N_1}\cdots q_D^{N_D}}{q^2}$ ($N_i$ even) is formally proportional to $\int dq \frac{q^N}{q^2}=0$, with $N=\sum_i N_i$ (using the standard  parametrization for $1/q^2=\int_0^\infty d\lambda \exp\left(-\lambda q^2\right)$). Therefore, in that case we have simply:

\begin{equation}
J_n(s;T=0)=\left(\frac{s}{2}\right)^n J(s),
\label{JnT0}
\end{equation}
with $J(s)$  given in \eqref{Js}. 

\begin{figure}
\centering
\includegraphics[scale=1]{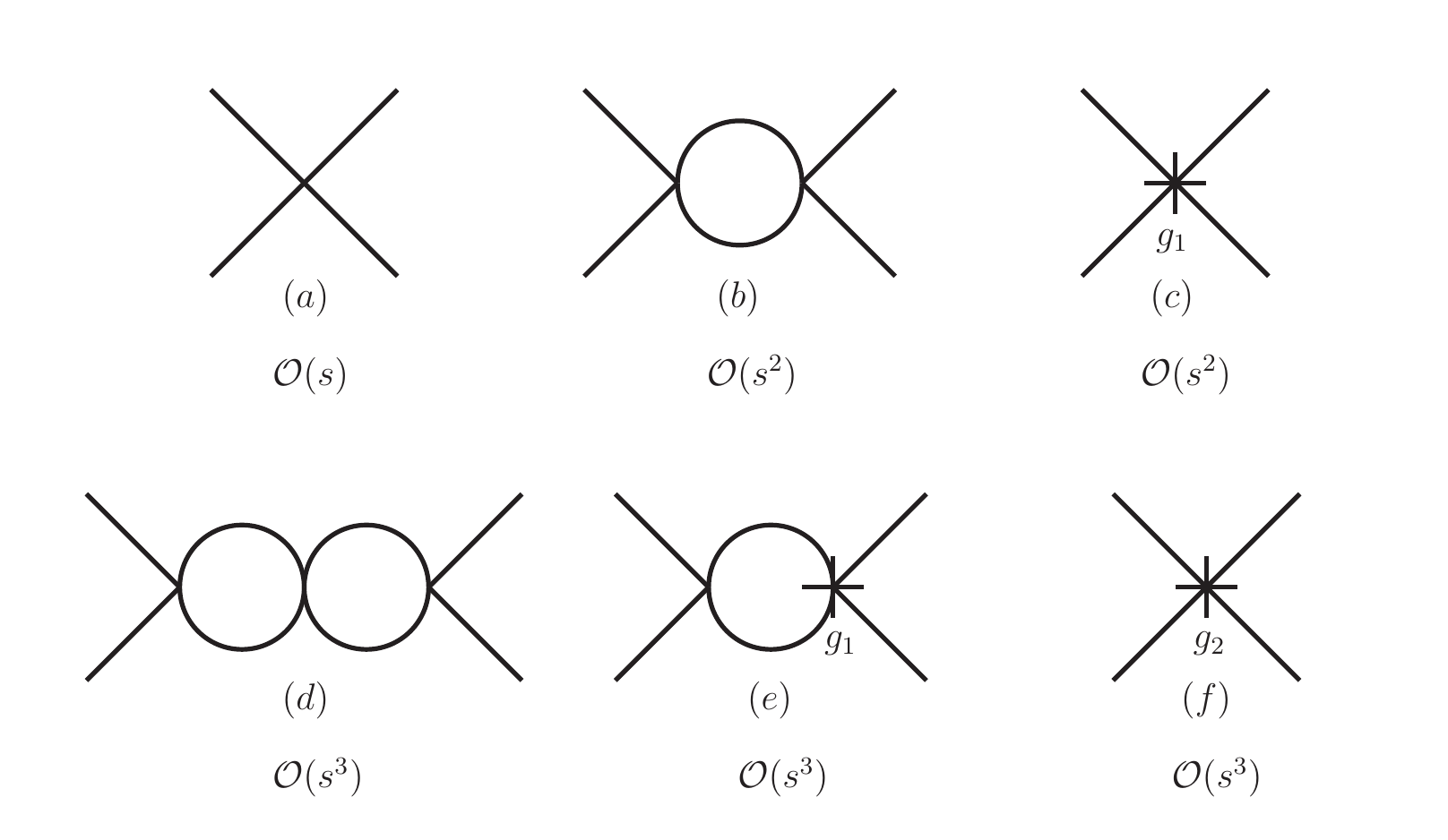}
\caption{Diagrams up to $\Od(s^3)$ for the renormalized amplitude at $T=0$. We plot the different topological configurations contributing, so that diagram (e) is multiplied by two, corresponding to the  possible vertex insertions of $g_1$.}
\label{fig:T0diags3}
\end{figure}

Thus, at $T=0$, any $g_k$ insertion is proportional to $s^{k+1}$, regardless of the vertex being internal of external.  As stated in the main text, this allows to renormalize the amplitude at every order. As an example, let us  show here the diagrams contributing up to $\Od(s^3)$ at $T=0$, which are those showed in Fig.\ref{fig:T0diags3} and where we have indicated the order of every diagram. Summing up these contributions, the amplitude is finite with the following renormalization of $g_1$, $g_2$:
\begin{align}
g^R_1(\mu)&=g_1+\frac{1}{2}J_\epsilon(\mu),
\label{reng1}\\
g^R_2(\mu)&=g_2-\frac{1}{4}\left[J_\epsilon(\mu)\right]^2+g^R_1(\mu)J_\epsilon(\mu)=g_2+\frac{1}{4}\left[J_\epsilon(\mu)\right]^2+g_1J_\epsilon(\mu).
\label{reng2}
\end{align}

When the $T=0$ amplitude is written in terms of the renormalized constants, it adopts the form (\ref{renormamp1}), where to this order $G_{R}(s;\mu)=1+g^{R}_{1}(\mu)\frac{s}{F^2}+g^{R}_{2}(\mu)\frac{s^2}{F^4}+\Od(s^3)$. As it was assured above, this is equivalent to renormalize the amplitude by the functional renormalization of the four-pion vertex given in eqns. \eqref{4piren} and \eqref{renormG}.

At $T\neq 0$, there are additional complications that need to be analyzed in detail. First of all,  the simple relation \eqref{JnT0} for the integrals $J_n$ in \eqref{Jn} does no longer hold. Namely, for $n=1$ we get directly from \eqref{Jn}:

\begin{equation}
J_1(p;T)=\frac{1}{2}\left[sJ_0+2I_\beta\right],
\label{J1}
\end{equation}
with $I_\beta$ the tadpole integral in \eqref{tadpole}, while for $n=2$,

\begin{equation}
J_2(p,T)=\frac{1}{4}\left[s^2J_0+4sI_\beta+\int_T d^Dq \frac{(p+q)^2}{q^2}+\int_T d^D q \frac{(p-q)^2}{q^2}\right]=\frac{s}{4}\left[sJ_0+2I_\beta\right],
\label{J2}
\end{equation}
where we have used that in DR  $\int_T d^Dq \ q^{2n}=\int_T d^Dq \ (p-q)^{2n}=0$ for $n=0,1,2,\dots$ (although not for any real $n$  as it happened in the $T=0$ case) since $\int d^{D-1} \ \vec{q} \modq^{2k}=0$ for  $k=0,1,2,\dots$. Note also that in \eqref{J2}, the two contributions $\int_T d^Dq \ \frac{p\cdot q}{q^2}$ cancel among them, and also independently by parity. However, this does not happen for $n\geq 3$, meaning that the $J_n$'s are not simply linear combinations of $J_0$ and $I_\beta$ as in the previous cases. For instance, for $n\geq 3$ the following integral contributes:

\begin{equation}
\int_T d^D q \frac{(p\cdot q)^2}{q^2}=p_\mu p_\nu I^{\mu\nu}(T)=-\omega_m^2 I^{00}(T)-\modp^2I_s(T),
\end{equation}
with $I^{\mu\nu}=\int_T d^D q \frac{q^\mu q^\nu}{q^2}$ and $I_s=\frac{1}{D-1}I^{j}_{\,j}$. At $T=0$, one has $I^{\mu\nu}=g^{\mu\nu}\frac{1}{D}\int d^D q=0$ in DR, but at $T\neq 0$ the timelike and spacelike contributions decouple, from the loss of Lorentz covariance in the thermal bath, and they are in general nonzero. Besides,

\begin{align}
I_s(T)=\frac{1}{D-1}\int_T d^Dq \modq^2 \int_0^\infty d\lambda \ e^{-\lambda (\omega_n^2+\modq^2)}=\frac{1}{(4\pi)^{\frac{D-1}{2}}}\frac{T}{2}\sum_{n=-\infty}^\infty  \int_0^\infty d\lambda \ e^{-\lambda \omega_n^2} \lambda^{-1-\frac{D-1}{2}}\nonumber\\
=I_s(0)+\frac{1}{(4\pi)^{\frac{D}{2}}} \sum_{k=1}^\infty  \int_0^\infty d\lambda \ \lambda^{-1-\frac{D}{2}} e^{-\frac{k^2}{4T^2\lambda}}=\frac{1}{2}g_0(0,T)=\frac{\pi^2}{90}T^4,
\label{Is}
   \end{align}
where we have made use of the standard Feynman  parametrization as well as Poisson's summation formula $\sum_n F(n)=\sum_k \int_{-\infty}^\infty dx F(x) \exp{(2\pi ikx)}$. The function $g_0(M,T)$ is defined in  \cite{Gerber:1988tt}. On the other hand, using again the DR properties, $I^{00}(T)=-(D-1)I_s (T)$. Note that $I_s(0)=0$ so that these pure thermal contributions would not give rise to new type of divergences, i.e. different from those coming from the standard loop integral in \eqref{JTsepar}.  

Therefore, the Feynman rules at $T\neq 0$ for $g_k$ insertions change with respect to the $T=0$ ones. Namely, a $g_k$ insertion in the generic diagram of Fig.\ref{fig:insertions} produces an integral of the type \eqref{Jn} for the internal loop momenta $q$ and  then is not equivalent to a simple $s^k$ power as for $T=0$. 

\begin{figure}
\centering
\includegraphics[scale=0.85]{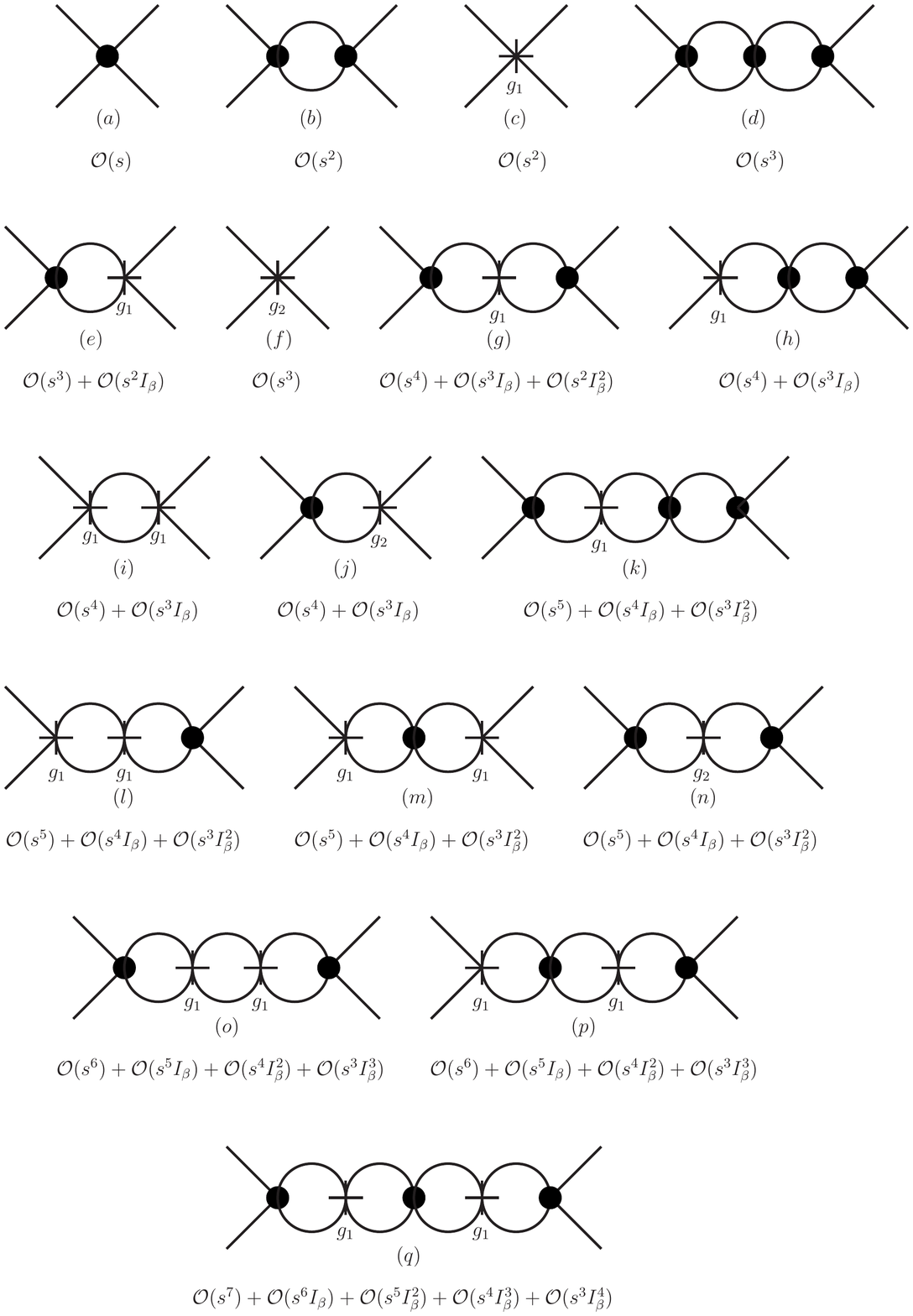}
\caption{Diagrams up to $\Od(s^3)$ for the renormalized amplitude at $T\neq0$. We plot the different topological configurations contributing, so that diagrams (e), (h), (j), (k), (l), (p) are multiplied by two, corresponding to the possible vertex insertions.}
\label{fig:finiteTdiags3}
\end{figure}

One of the consequences of the above results is that when considering all the diagrams contributing to a given $s^k$ order, the $s$ and $T^2$ powers mix, so that a larger number of diagrams has to be considered. In Fig.\ref{fig:finiteTdiags3} we have displayed all the diagrams that would give  $\Od(s^3)$ contributions, all of them including $g_1$ and $g_2$ insertions according to the results \eqref{J1} and \eqref{J2}. The vertex with no $g_{1,2}$ insertions is the effective thermal vertex in Fig.\ref{EffVertx}. Attached to each diagram, we have indicated the different powers of $s^nI_\beta^m$ that it gives rise to. 

We have calculated all diagrams in Fig.\ref{fig:finiteTdiags3} with the Feynman rules discussed above. The result is that the amplitude to that order remains finite with the same $T=0$ renormalizations of $g_1$ and $g_2$ given in \eqref{reng1}-\eqref{reng2}, which is a nontrivial consistency check. Furthermore, the analysis of the result reveals some interesting features that will shed light on the renormalization scheme to be followed in the general case. 

First of all, we show the $\Od(s^2)$ calculation. In addition to diagrams (a), (b), (c) of Fig.\ref{fig:finiteTdiags3}, which are the counterparts of (a), (b), (c) in Fig.\ref{fig:T0diags3} with the thermal vertex, we have to consider as well  diagram (e) in Fig.\ref{fig:finiteTdiags3}, since it includes a $I_\beta s^2$ contribution, as well as diagram (g), whose $I_\beta^2 s^2$ part has  to be taken into account. Altogether, we obtain for the amplitude at that order:

\begin{align}
A(p;T)&=\frac{s}{NF^2}f(I_\beta)\left\{1+\frac{s}{2F^2}f(I_\beta)\left[2g_1+J(p;T)\right]\right\}+\Od(s^3)\nonumber \\
&=\frac{s}{NF^2}f(I_\beta)\left\{1+\frac{s}{2F^2}f(I_\beta)\left[2g_1^R(\mu)+J_{fin}(p;T;\mu)\right]\right\}+\Od(s^3),
\label{ampTs2}
\end{align}
where we have separated the loop integral into its divergent and finite parts according to \eqref{JTsepar} and \eqref{Jepsilon} and we have used exactly the same renormalization of the $g_1$ constant as for $T=0$, namely \eqref{reng1}. The thermal amplitude in \eqref{ampTs2} is explicitly finite and scale independent to this order. Moreover, note that  we can write the thermal amplitude to that order in a  form similar to the renormalized $T=0$ case in \eqref{renormamp0} as follows:

\begin{align}
A(p;T)=\frac{s G_0(s)}{NF^2}\frac{1}{1-G_0(s)I_\beta/F^2}\frac{1}{1-\frac{s G_0(s)}{2F^2}\frac{1}{1-G_0(s)I_\beta/F^2}J(p;T)}+\Od(s^3)\nonumber\\=
\frac{s G_0(s)f[G_0(s) I_\beta]}{NF^2}\frac{1}{1-\frac{s G_0(s)f[G_0(s) I_\beta]}{2F^2}J(p;T)}+\Od(s^3),
\end{align}
with $G_0(s)$ given by the same expression as in the $T=0$ analysis \eqref{4piren}  to this order, i.e., $G_0(s)=1+g_1(s/F^2) +\Od(s^2)$. Thus,

\begin{align}
\frac{1}{A(p;T)}=&\frac{NF^2}{s}\left[\frac{1}{G_0(s)f[G_0(s) I_\beta]}-\frac{s J(p;T)}{2F^2}\right]=\frac{NF^2}{s}\left[\frac{1}{G_0(s)}-\frac{I_\beta}{F^2}-\frac{s J(p;T)}{2F^2}\right]
\nonumber\\
=&\frac{NF^2}{s}\left[\frac{1}{G_R(s;\mu)}-\frac{I_\beta}{F^2}-\frac{s J_{fin}(p;T;\mu)}{2F^2}\right]\nonumber\\
\Rightarrow& A_R(p;T)=\frac{s G_R(s;\mu)f[G_R(s;\mu) I_\beta]}{NF^2}\frac{1}{1-\frac{s G_R(s;\mu)f[G_R(s;\mu) I_\beta]}{2F^2}J_{fin}(p;T;\mu)}+\Od(s^3).
\end{align}
Here $G_R$ is written in terms of $G_{0}$ as in \eqref{renormG}, so that $G_{R}(s)=1+g_{1}^{R}(\mu)(s/F^2) +\Od(s^2)$, which renders the amplitude finite. 

The same structure is obtained when we calculate up to $\Od(s^3)$ and then we take into account the corresponding contributions from the diagrams in Fig.\ref{fig:finiteTdiags3}. Now we obtain

\begin{align}
A(p;T)=&\frac{s}{NF^2}f(I_\beta)\left\{1+\frac{s}{2F^2}f(I_\beta)\left[2g_1+J(p;T)\right]\right.
\nonumber\\+&\left.\frac{s^2}{F^4}f^2(I_\beta)\left[g_1J(p;T)+\frac{1}{4}J^2(p;T)+g_2\left(1-\frac{I_\beta}{F^2}\right)+g_1^2\frac{I_\beta}{F^2}\right]\right\}+\Od(s^4)\nonumber \\=&
\frac{s}{NF^2}f(I_\beta)\left\{1+\frac{s}{2F^2}f(I_\beta)\left[2g_1^R(\mu)+J_{fin}(p;T;\mu)\right]\right.\nonumber\\
+&\left.\frac{s^2}{F^4}f^2(I_\beta)\left[g^R_1 (\mu)J_{fin}(p;T;\mu)+\frac{1}{4}J_{fin}^2(p;T;\mu)+g_2^R(\mu)\left(1-\frac{I_\beta}{F^2}\right)+\left[g_1^R(\mu)\right]^2\frac{I_\beta}{F^2}\right]\right\}+\Od(s^4),
\label{ampTs3}
\end{align}
which is again finite with the renormalizations \eqref{reng1} and \eqref{reng2} and can also be written as

\begin{align}
A(p;T)=&\frac{s G_0(s)f[G_0(s) I_\beta]}{NF^2}\frac{1}{1-\frac{s G_0(s)f[G_0(s) I_\beta]}{2F^2}J(p;T)}+\Od(s^4)
\nonumber\\\Rightarrow& A_R(p;T)=\frac{s G_R(s;\mu)f[G_R(s;\mu) I_\beta]}{NF^2}\frac{1}{1-\frac{s G_R(s;\mu)f[G_R(s;\mu) I_\beta]}{2F^2}J_{fin}(p;T;\mu)}+\Od(s^4),
\end{align}
with $G_{0}(s)=1+g_{1}(s/F^2)+g_{2}(s/F^2)^2+\Od(s^3)$ and $G_{R}(s;\mu)=1+g_{1}^{R}(\mu)(s/F^2)+g_{2}^{R}(\mu)(s/F^2)^2+\Od(s^3)$.

Therefore, from the previous expressions we observe that the $T\neq 0$ renormalization is equivalent to the following renormalization of the four-pion thermal effective vertex:

\begin{equation}
\frac{s}{NF^2}f(I_\beta)\rightarrow \frac{s}{NF^2} G_0(s) f\left[G_0(s) I_\beta\right].
\label{4effren}
\end{equation}

What is interesting for our purposes is that the renormalization given in \eqref{4effren} can actually be achieved by a $T=0$ renormalization of each of the $2n$-pion vertices in the original Lagrangian by assigning them a $G_0^{n-1}(s)$ factor in momentum space, as displayed in Table \ref{feynruletable}, thus generalizing the $4\pif$ vertex renormalization in \eqref{4piren}.

\begin{table}
\begin{tabular}{cccc}  \hline \hline
	Vertex & Lagrangian & Bare rule & Renormalized rule \\ \hline
	4 pions & $\displaystyle -\frac{\pif^2\square\pif^2}{8NF^2}$ & $\displaystyle\frac{s}{NF^2}$ & $\displaystyle\frac{sG_{0}(s)}{NF^2}$  \\ \\
	6 pions & $\displaystyle-\frac{\left(\pif^2\right)^2\square\pif^2}{16 (NF^{2})^{2}}$ & $\displaystyle\frac{s}{(NF^{2})^{2}} I_\beta$ & $\displaystyle\frac{sG_{0}^{\,2}(s)}{(NF^{2})^{2}} I_\beta$ \\ \\
	$2k+4$ pions & $\displaystyle -\frac{\left(\pif^2\right)^{k+1}\square\pif^2}{8(k+1)(NF^{2})^{k+1}}$ & $\displaystyle\frac{s}{(NF^{2})^{k+1}} I_\beta^k$ & $\displaystyle\frac{sG_{0}^{\,k+1}(s)}{(NF^{2})^{k+1}}I_\beta^k$ \\
	\hline \hline
\end{tabular}
\caption{Feynman rules renormalization for the interaction vertices of the Lagrangian \eqref{NLSM} for $\pi\pi$ scattering at large $N$.}
\label{feynruletable}
\end{table}

It is actually possible to trace  the origin of this renormalization scheme in terms of the contributing diagrams. Consider for instance diagrams with just one $g_1$ insertion. At $\Od(s^2)$, one has to sum the contributions to the amplitude from diagrams (c), (e) and (g) in Fig.\ref{fig:finiteTdiags3}, namely,

\begin{equation}
\frac{g_1 s^2}{NF^4}\left[1+\frac{2x}{1-x}+\frac{x^2}{(1-x)^2}\right]=\frac{g_1 s^2}{NF^4}\sum_{k=1}^\infty k x^{k-1},
\label{renint1g1}
\end{equation}
with $x=I_\beta/F^2$. Now,  the infinite contributions in the sum in \eqref{renint1g1} amount to four-point diagrams with $k-1$ tadpole contractions ($k=1,2,\dots$) i.e., like the diagrams in Fig.\ref{EffVertx} with a $kg_1 \frac{s^2}{NF^4}$ vertex. But we can interpret each of those diagrams as the contribution to the scattering amplitude of a multiplicative renormalization of the $(2+2k)$-pion vertex of the Lagrangian \eqref{NLSM} given by $\frac{s}{(NF^2)^k}\rightarrow \frac{s}{(NF^2)^k} \left[1+kg_1\frac{s}{F^2}\right]+\Od(s^3)=\frac{s}{(NF^2)^k} G_0^k(s)+\Od(s^3)$, for which closing $2(k-1)$ lines gives $(NI_\beta)^{k-1}$ to leading order in $N$. This is precisely the rule given in Table \ref{feynruletable}. 

We have checked that we find the same rule by analyzing  the remaining  $g_i$ insertions in  the graphs in Fig.\ref{fig:finiteTdiags3}. Specifically, the $\Od(s^3)$ contribution with one $g_1$ insertion (diagrams (e), (g), (h) and (k) ) give the renormalization rule in Table \ref{feynruletable} for diagrams with the one-loop $J$ function, the $\Od(s^3)$ with one $g_2$ insertion (diagrams (j) and (n)) give the linear part in $g_2$ of the  $G_{0}^{k}(s)$ contribution which we have just analyzed above to $\Od(s^2)$, while the $\Od(s^3)$ with two $g_1$ insertions (diagrams (i), (l), (m), (o), (p), (q)) reproduce precisely the $g_{1}^{2}$ part of those $G_0^k(s)$ terms.  Therefore,  in this way we are able to reinterpret all $g_{i}$ insertions in Fig.\ref{fig:finiteTdiags3} giving  $s^nI_\beta^m$ mixed powers with $n=2,3$ and $m=1,2,3,4$, in terms of the $T$-independent renormalization scheme in Table \ref{feynruletable} and \eqref{4effren}, from the contributing diagrams without mixed terms. For higher insertions, we would need higher order diagrams with respect to those of Fig. \ref{fig:finiteTdiags3},  i.e., up to $\Od(s^4)$.

The crucial conclusion is that following the above renormalization scheme also in the general nonperturbative case, namely starting from the full amplitude \eqref{thermalampbare}, yields a finite scattering amplitude with a $T=0$  renormalization, as discussed in the main text. 

Finally, let us comment about this renormalization scheme from the point of view of the effective Lagrangian. For that purpose, Let us write down the expansion of the NLSM Lagrangian (\ref{NLSM}) as

\begin{align}
\mathcal{L}&=\frac{1}{2}g_{ab}(\pif)\partial_{\mu}\pif^{a}\partial^{\mu}\pif^{b}=\frac{1}{2}\left(\delta_{ab}+\frac{1}{NF^{2}}\frac{\pif_{a}\pif_{b}}{1-\pif^2/NF^{2}}\right)\partial_{\mu}\pif^{a}\partial^{\mu}\pif^{b} \notag \\
&=\frac{1}{2}\left[\partial_{\mu}\pif_{a}\partial^{\mu}\pif^{a}+\frac{1}{4}\sum_{k=0}^{\infty}{\frac{(\pif^2)^{k}(\partial_{\mu}\pif^2)(\partial^{\mu}\pif^2)}{(NF^{2})^{k+1}}}\right]
\notag \\
&=-\frac{1}{4}\left[1+\frac{1}{2}\sum_{k=0}^{\infty}{\frac{1}{k+1}\left(\frac{\pif^2}{NF^{2}}\right)^{k+1}}\right]\square\pif^2=-\frac{1}{4}\left[1-\frac{1}{2}\ln\left(1-\frac{\pif^2}{NF^{2}}\right)\right]\square\pif^2,
\label{laglogform}
\end{align}
where we have used

\begin{align*}
(\pif^2)^{j}(\partial_{\mu}\pif^2)(\partial^{\mu}\pif^2)&=\frac{1}{j+1}\left\{\partial_{\mu}\left[(\pif^2)^{j+1}\partial^{\mu}(\pif^2)\right]-(\pif^2)^{j+1}\square\pif^2\right\}; \\
\partial_{\mu}\pif_{a}\partial^{\mu}\pif^{a}&=\partial_{\mu}(\pif_{a}\partial^{\mu}\pif^{a})-\frac{1}{2}\square\pif^2,
\end{align*}
as well as integration by parts.

From the previous expression,  we see that the Feynman rules for the renormalized Lagrangian listed in Table   \ref{feynruletable}, as long as $\pi\pi$ scattering in the large-$N$ limit is concerned, are equivalent to replace:

\begin{equation}
 (\pif^2)^{k+1}\square\pif^2\rightarrow (\pif^2)^{k+1}G_0^{k+1}(-\square)\square\pif^2.
 \label{replag}
 \end{equation}

Therefore, the renormalization scheme we are analyzing here is very natural in the sense that every $\pif^2$ power in the expansion is renormalized with the same power of the $G_0$ function, acting as showed in  \eqref{replag}. Since those powers come from the same metric covariant function, as indicated in \eqref{laglogform}, this is consistent with the introduction of the counterterm Lagrangians in a covariant fashion as we have discussed in section \ref{sec:ren}.

\end{document}